\documentclass[12pt]{iopart}
 
\usepackage{iopams}
\expandafter\let\csname equation*\endcsname\relax
\expandafter\let\csname endequation*\endcsname\relax
\usepackage{amsmath}
\usepackage{graphicx}

\usepackage[width=\linewidth,textfont=small]{caption}
\usepackage{relsize}

\usepackage{bigstrut}
\usepackage{multirow}

\begin{document}

\title{Single-crystal graphene on Ir(110)}

\author{Stefan Kraus$^1$, Felix Huttmann$^1$, Jeison Fischer$^1$, Timo Knispel$^1$ , Ken Bischof$^1$, Alexander Herman$^2$, Marco Bianchi$^3$, Raluca-Maria Stan$^3$, Ann Julie Holt$^3$, Vasile Caciuc$^4$, Shigeru Tsukamoto$^4$, Heiko Wende$^2$, Philip Hofmann$^3$, Nicolae Atodiresei$^4$, and Thomas Michely$^1$}
\address{$^1$ II. Physikalisches Institut, Universität zu K\"{o}ln, Zülpicher Str. 77, 50937 K\"{o}ln, Germany}
\address{$^2$ Faculty of Physics and Center for Nanointegration Duisburg-Essen (CENIDE), University of Duisburg-Essen, Lotharstraße 1, 47048 Duisburg, Germany}
\address{$^3$ Department of Physics and Astronomy, Interdisciplinary Nanoscience Center (iNANO), Aarhus University, 8000 Aarhus C, Denmark}
\address{$^4$ Peter Gr{\"{u}}nberg Institute and Institute for Advanced Simulation, Forschungszentrum Jülich, Wilhelm-Johnen-Straße, 52428 J{\"{u}}lich, Germany}
\ead{kraus@ph2.uni-koeln.de (experiment); n.atodiresei@fz-juelich.de (theory)}

\vspace{10pt}
\begin{indented}
\item[]\today
\end{indented}


\begin{abstract}
A single-crystal sheet of graphene is synthesized on the low-symmetry substrate Ir(110) by thermal decomposition of C$_2$H$_4$ at 1500\,K. Using scanning tunneling microscopy, low-energy electron diffraction, angle-resolved photoemission spectroscopy, and \textit{ab initio} density functional theory the structure and electronic properties of the adsorbed graphene sheet and its moiré with the substrate are uncovered. The adsorbed graphene layer forms a wave pattern of nm wave length with a corresponding modulation of its electronic properties. This wave pattern is demonstrated to enable the templated adsorption of aromatic molecules and the uniaxial growth of organometallic wires. Not limited to this, graphene on Ir(110) is also a versatile substrate for 2D-layer growth and makes it possible to grow epitaxial layers on ureconstructed Ir(110).
\end{abstract}

%

%
%
\maketitle
 

\section{Introduction}

\textit{In-situ} grown graphene (Gr) is an excellent inert substrate for subsequent growth and van der Waals epitaxy. Examples are the growth of EuO(001) on Gr/Ir(111)~\cite{Klinkhammer2013}, or the van der Waals epitaxy of transition metal dichalcogenide layers on bilayer Gr/6H-SiC(0001)~\cite{Ugeda2014,Ugeda2016} or on Gr/Ir(111)~\cite{Hall2018,Murray2019}. Due to its inertness, Gr is also well suited as substrate when investigating the properties of molecular layers in the absence of strong molecule-substrate hybridization \cite{Garnica2013, Hamalaien2012}. Similarly, its inertness makes it an ideal substrate for organometallic chemistry with the function to confine the reactand diffusion to two dimensions \cite{Huttmann2017}. In case of Gr forming a moiré with its growth substrate, also templating of molecular layers \cite{Mao2009} and of atom or cluster superlattices has been reported~\cite{Baltic2016, Hartl2020}.

Gr growth also has a substantial effect on its substrate: step edges are moved \cite{Borovikov2009}, step bunches and facets are formed~\cite{Kraus13,Bao2016}, or vicinal growth substrates become faceted~\cite{Srut2016}. A layer of Gr was shown to protect a surface against oxidation, e.g. for Ni(111)~\cite{Dedkov2008} or Pt(100)~\cite{Nilsson2012}, or to prevent the lifting of a reconstruction \cite{Nilsson2012b}. Noteworthy, the protection of a metal surface against the formation of a surface reconstruction has not yet been reported.
\\
Up to now mostly symmetry-matching substrates were used for \textit{in-situ} growth of Gr, e.g. fcc(111) or hcp(0001) surfaces~\cite{Batzill2012}. Depending on the strength of interaction between Gr and the substrate, either single-domain Gr could be grown, e. g. for Ru(0001)~\cite{Marchini2007} (good orientation due to strong interaction) or multidomain structures result, e.g. for Pt(111) \cite{Land1992} or Cu(111)~\cite{Gao2010}. For strongly interacting substrates forming a moiré with Gr the corrugation is often substantial due to the spatial variation of binding within the moiré unit cell \cite{Martoccia2010}. Gr/Ir(111) is a unique case, in which the interaction is still weak, but due to proper selection of growth conditions, a well-oriented single-crystal Gr sheet can still be grown \cite{Loginova2009,Hattab2011}.
\\
Far less work has been conducted for Gr growth on non-symmetry-matching substrates of fourfold \cite{Nilsson2012} and twofold symmetry \cite{Fedorov2011,Dugerjav2020,Achilli2018,Vinogradov2012,Dai2016}. Gr on fcc(110) metal surfaces displays domain formation or multiple orientations. This holds for Ni(110), Cu(110) and 
Pt(110), irrespective of whether the interaction with the substrate is strong or weak \cite{Fedorov2011,Dugerjav2020,Achilli2018}. Up to now, among the metals only the quasihexagonal dense-packed bcc(110) surface of Fe, still only of twofold symmetry, was shown to enable Gr growth with unique orientation \cite{Vinogradov2012}. Intense research was triggered by the finding that on the (110) face of the semiconductor Ge  growth of large Gr single-crystal layers is 
possible~\cite{Lee2014,Dai2016}, though at the risk of growth close to substrate melting. Also for the isostructural and isoelectric hexagonal boron nitride (h-BN) monolayers, the growth of single-domain phase-pure layers on only twofold-symmetric substrates was not feasible on metals \cite{Corso2005,Greber2006,Allan2007,Martinez-Galera2018}. The exception is the growth of h-BN on Pt(110), where the adlayer imposes a complex reconstruction change of the substrate 
\cite{Steiner2019}. 
\\
Here, we introduce Gr on the low-symmetry substrate Ir(110), which displays single-domain single-crystal growth when choosing the proper growth conditions. The perfection by which the Gr layer can be fabricated is surprising, when considering that the clean Ir(110) surface is heavily reconstructed at room temperature. Unreconstructed Ir(110) forms a ridge pattern of $(331)$ and $(33\bar{1})$ nano-facets with a corrugation in the nm range \cite{Koch1991}. Under the Gr cover, however, Ir(110) remains unreconstructed. 
\\
Using scanning tunneling microscopy (STM) and low-energy electron diffraction (LEED) the moiré of Gr with Ir(110) is determined. Density functional theory (DFT) calculations reveal a strong modulation of binding and charge transfer to Gr associated with the moiré wave pattern along the $[1\bar{1}0]$ direction of the substrate with a periodicity of $10$\,\AA.
\\
The use of this wave pattern for templated adsorption is directly demonstrated here. In DFT calcuations naphthalene is used as a paradigm for an aromatic molecule to explore the anisotropic energy landscape of physisorption induced by the wave pattern. Based on these insights, unaxial alignment of sandwich-molecular wires during organometallic on-surface synthesis \cite{Huttmann2017,Huttmann2019} is experimentally demonstrated.
\\
Inert single-crystal substrates are rare, but attractive for the growth of quasi-freestanding 2D layers. The application potential of Gr/Ir(110) for 2D-layer growth is exemplified through molecular beam epitaxy of monolayer NbS$_2$. As a last example of versatility, we show that through Gr intercalation thermodynamically stable epitaxial layers can be grown on \textit{unreconstructed} Ir(110).

\section{Methods}
Gr on Ir(110) was synthesized with identical results on two different crystals in the ultrahigh vacuum systems ATHENE (base pressure below $ 1 \cdot 10^{-10}$\,mbar, STM imaging at $300$\,K) and M-STM (base pressure in preparation chamber $ 3 \cdot 10^{-10}$\,mbar, in STM chamber below 
$ 1 \cdot 10^{-11}$\,mbar, STM imaging temperature $1.7$\,K). Gases are delivered through a gas dosing tube giving rise to a pressure enhancement by a factor of 80 compared to the pressure measured through a distant ion gauge specified here. Sample cleaning was accomplished by exposure to $ 1 \cdot 10^{-7}$\,mbar oxygen at $1200$\,K when needed, cycles of noble gas sputtering (Ar, Xe), and brief annealing to $1500$\,K. Closed layers of Gr on Ir(110) were grown by exposure to $ 2 \cdot 10^{-7}$\,mbar ethylene for $210$\,s at $1500$\,K for the single-domain Gr phase. For the two-domain Gr phase briefly mentioned in the manuscript and discussed in the SI, the ethylene exposure was at $1300$\,K. EuCot sandwich-molecular wires were grown at a sample temperature of 300\,K by sublimation of Eu from a Knudsen cell with a deposition rate of $1.1 \cdot 10^{17} \, \frac{\mathrm{atoms}}{\mathrm{m}^2\mathrm{s}}$ in a background pressure of $1 \cdot 10^{-8}$\,mbar Cot \cite{Huttmann2017}. The software \emph{WSxM}\cite{Horcas2007} was used for STM data processing.
\\
To ensure Gr quality and for structural characterization LEED was used in an energy range of $30-150$\,eV. The LEED patterns shown are contrast-inverted for better visibility.
\\
Angle-resolved photoemission spectroscopy (ARPES) measurements have been conducted at the SGM-3 beamline at the synchroton ASTRID2 in Aarhus, Denmark. The samples were grown in situ in an ultrahigh vacuum chamber (base pressure $3 \cdot 10^{-10}$\,mbar) connected to the beamline, and using the recipe described above. Sample cleaning has been accomplished by noble gas sputtering using Ne. The samples have been checked in situ using the Aarhus STM mounted at the endstation to check for consistency with the homelab results.
\\
Our \textit{ab initio} density functional theory (DFT) \cite{PR136_B864,PR140_A1133} were carried 
out using the projector augmented wave method (PAW) \cite{PRB50_17953} as implemented in the VASP code \cite{PRB47_558,PRB54_11169,PRB59_1758}. 
The van der Waals interactions present in the Gr/Ir(110) 
system were taken into account by employing the non-local correlation 
energy functional vdW-DF2 \cite{PRB82_081101R} together with a 
re-optimized \cite{PRB89_121103R} Becke (B86b) exchange energy 
functional \cite{JCP86_7184}. 
Gr on Ir(110) was modeled by a slab containing three Ir  
layers and a vacuum region of $\approx 21$\,\AA\: amounting to 350 C and 264 Ir atoms, respectively. 
The ground-state geometry and its electronic structure of this system 
have been obtained for a kinetic energy cut-off of $500$\,eV and a threshold 
value of the calculated Hellmann-Feynman forces of $\approx 0.005$\,eV$/$\AA. 
Furthermore, for the structural relaxation the Brillouin zone integrations 
were performed using the $\Gamma$ point while the density of states (DOS) 
was obtained with the help of a 2$\times$2 $k$ mesh. 

\section{Results and discussion}

\begin{figure}[!ht]
\centering
\includegraphics[width=12cm]{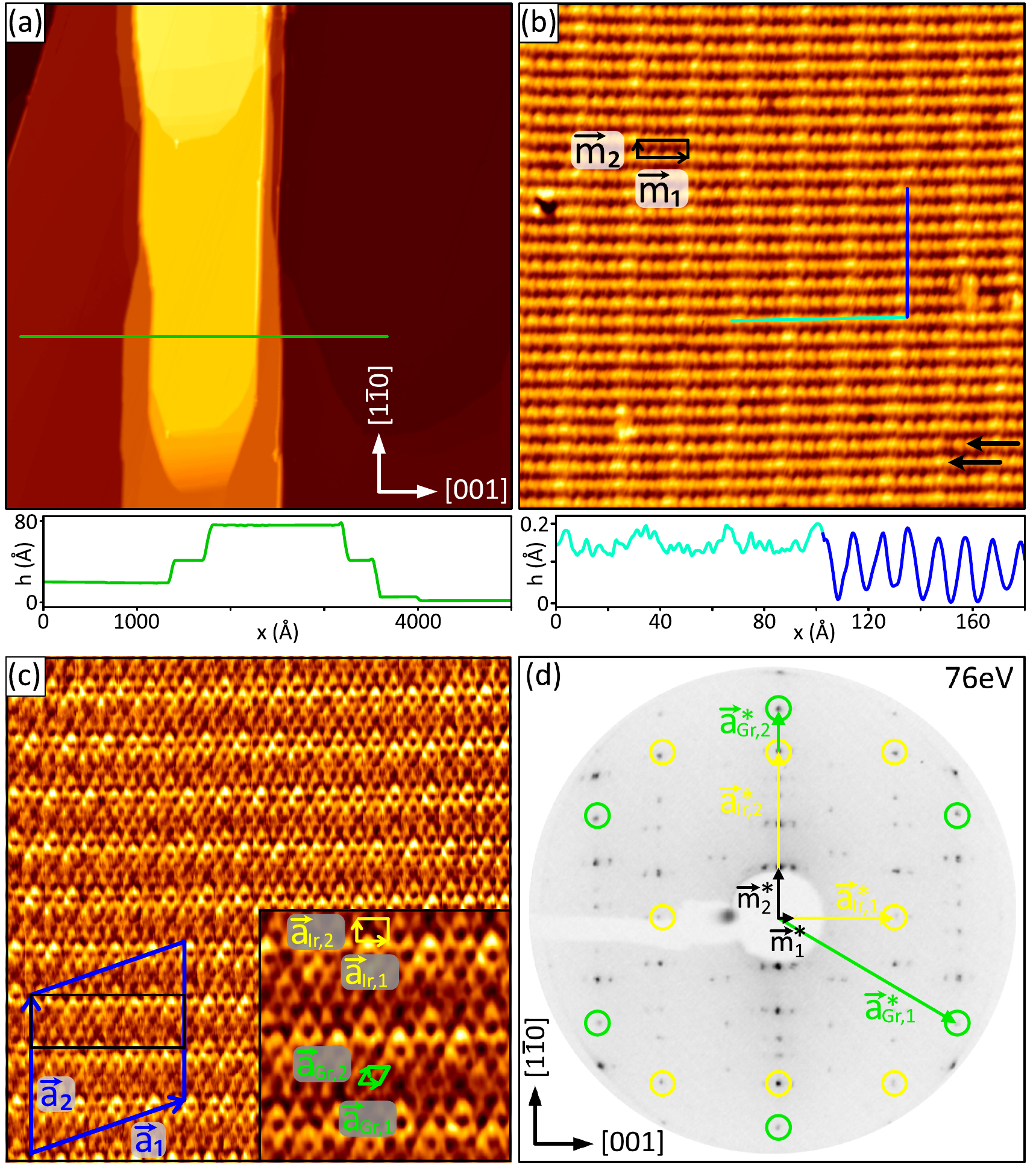}
\caption{(a) Large scale STM topograph ($7000 \times 7000$\,\AA$^2$) of Gr/Ir(110) grown at $T=1500$\,K. The $\left[001\right]$ and $\left[1\bar{1}0\right]$ directions specified in the lower right corner are valid also for (b), (c) and (d). STM height profile along the green line is shown below the topograph. (b) STM topograph ($300 \times 300$\,\AA$^2$) of flat terrace area. Black rectangle indicates the moiré unit cell. Black arrows highlight varying appearance of wave crests. STM height profile along path of cyan and blue line is shown below topograph. (c) Atomic resolution STM topograph ($100 \times 100$\,\AA$^2$) with commensurate superstructure cell and primitive translations $\vec{a}_1$ and $\vec{a}_2$ indicated in blue. Moiré unit cell is shown in black. Inset: magnified view ($30 \times 30$\,\AA$^2$) with centers of Gr rings visible as dark depressions. Gr and Ir(110) unit cells and primitive translations are indicated. (d) $76$\,eV LEED pattern of Gr/Ir(110). First order Ir and Gr reflections are encircled yellow and green, respectively. Reciprocal moiré, Gr and Ir primitive translations are indicated. See text. STM imaging temperatures are $300$\,K in (a) and (b), and $1.7$\,K in (c). Tunneling parameters are (a) $U_{\mathrm{bias}}=-1.33$\,V and $I_{\mathrm{t}}=0.8$\,nA, (b) $U_{\mathrm{bias}}=-1.33$\,V and $I_{\mathrm{t}}=1.1$\,nA, (c) $U_{\mathrm{bias}}=-0.05$\,V and $I_{\mathrm{t}}=20.0$\,nA.}
\label{fig:structure}
\end{figure}

\textbf{Superstructure -- experiment:} Gr is grown on carefully cleaned Ir(110) through exposure to ethylene at $1500$\,K. After cooldown to $300$\,K, STM finds large flat terraces, separated by plateaus with flat top levels and elongated along the $[1\bar{1}0]$ direction. The profile in Figure~\ref{fig:structure}a shows a plateau height of several nm. The plateau top level and the surrounding base level are flat. There is no indication for a ridge pattern of $(331)$ and $(33\bar{1})$ nano-facets with a corrugation in the nm~range \cite{Koch1991}, as it is observed after cooldown in the absence of a Gr cover [compare Figure~S1 of the Supporting Information (SI)]. A zoom into a flat terrace area makes a twofold symmetric moiré visible, as shown in Figure~\ref{fig:structure}b. We define the rectangular moiré cell as well as the moiré vectors $\vec{m}_1$ and $\vec{m}_2$ as indicated in Figure~\ref{fig:structure}b. The moiré leads to a well visible wave pattern with wave vector in the direction of $\vec{m}_2$, i.e. along the $[1\bar{1}0]$ direction. The wave crests and troughs are consequently oriented along the $[001]$ direction. The additional, larger wavelength periodicity with wave vector in the direction of $\vec{m}_1$ is less pronounced. 

Both periodicities can be recognized in the profile of Figure~\ref{fig:structure}b. With the profiles taken on the wave pattern crests and under the tunneling conditions chosen, the corrugation is $0.08$\,\AA\: for the long wavelength periodicity with wave vector along $\vec{m}_1$ (cyan line in Figure~\ref{fig:structure}b), while it is $0.17$\,\AA\: for the wave pattern with wave vector along $\vec{m}_2$ (blue line in Figure~\ref{fig:structure}b). It can be seen in Figure~\ref{fig:structure}b that on some wave crests an additional corrugation with a wavelength of about $\frac{1}{4}$ of the moiré periodicity is present. By visually analyzing this extra corrugation along the wave crests highlighted by black arrows in Figure~\ref{fig:structure}b it is obvious that (i) this extra corrugation varies along the wave crests and that (ii) neighboring wave crests differ in the amplitude of this corrugation, being almost absent or quite pronounced. While (i) suggests the presence of small tilts in the orientation of the two lattices, (ii) indicates that the moiré is not commensurate along $\vec{m}_2$. The large wavelength periodicity with wave vector in the direction of $\vec{m}_1$ is also not perfectly uniform and affected by the presence of defects. This is apparent for the vertical crests at the right of Figure~\ref{fig:structure}b, where the crests slightly change orientation due to the defect in the middle right of the topograph. Again, this indicates the presence of small tilts and shears of the Gr lattice.

We note that depending on the tunneling parameters and tip condition the corrugation may differ substantially from the apparent corrugation found for the topograph of Figure~\ref{fig:structure}b. Corrugations of up to $0.4$\,\AA\: for the long wavelength periodicity with wave vector along $\vec{m}_1$ and up to $0.8$\,\AA\: for the wave pattern with wave vector along $\vec{m}_2$ are found. The corrugation along $\vec{m}_1$ is generally found to be smaller than the one along $\vec{m}_2$. Elastic effects of tip-surface interaction at very low tunneling resistances, and at somewhat larger resistances a distance dependence of the corrugation on the local density of states are most likely the origin of this variation.

In the atomically resolved STM topograph of Figure~\ref{fig:structure}c, the centers of the Gr honeycombs are well visible as dark depressions. The Gr zigzag rows are close to perfectly aligned with the $[001]$ direction. The moiré periodicity $m_2$ along the $[1\bar{1}0]$ direction is well visible and caused by the superposition of the $2.715$\,\AA\: periodicity of Ir(110) and the zigzag row spacing of Gr. Application of the moiré construction for the case of aligned periodicities as outlined in~\cite{NDiaye2008} results in $m_2 = (9.94\pm 0.15)$\,\AA\: and a zigzag row spacing of $(2.133\pm0.007)$\,\AA. This row spacing implies a Gr lattice parameter of $a_{\mathrm{Gr}}=(2.463\pm0.008)$\,\AA. The Gr lattice parameter agrees within the limits of error with the in-plane lattice parameter of relaxed graphite $a_{\mathrm{graphite}}=2.4612$\,\AA~\cite{Eckerlin1971}. Gr on Ir(110) appears to be unstrained. The determination of the periodicity $m_1$ is more difficult, since it is not clearly visible in the atomically resolved topograph of Figure~\ref{fig:structure}c and similar ones, presumably due to the low tunneling resistance conditions needed to obtain atomic resolution. However, the observation of the ratio of $m_1$ to $m_2$ is possible in STM topographs without atomic resolution
and allows one to estimate $m_1 = (33\pm2)$\,\AA. Finally, the Fourier transform of Figure~\ref{fig:structure}c displays spots corresponding to the Ir(110) and Gr lattices simultaneously (compare Figure~S2 in the SI). Its analysis confirms the absence of significant strain in Gr and the magnitude of the Gr primitive translations $\vec{a}_{\mathrm{Gr},1}$ and $\vec{a}_{\mathrm{Gr},2}$ to deviate less than 0.5\,\%.

The LEED pattern of Gr/Ir(110) shown in  Figure~\ref{fig:structure}d can be decomposed into reflections of \textit{unreconstructed} Ir(110) encircled yellow and first order reflections of a single Gr domain encircled green. All other reflections are linear combinations of the Ir(110) and Gr reciprocal lattice vectors. The moiré periodicity along $[001]$ corresponds to the difference $\vec{m}_{1}^*=2\cdot\vec{a}_{\mathrm{Gr},1}^*+\vec{a}_{\mathrm{Gr},2}^*-3\cdot\vec{a}_{\mathrm{Ir},1}^*$ and along $[1\bar{1}0]$ to $\vec{m}_{2}^*=\vec{a}_{\mathrm{Gr},2}^*-\vec{a}_{\mathrm{Ir},2}^*$. The vectors $\vec{m}_{i}^*$, $\vec{a}_{\mathrm{Gr},i}^*$ and $\vec{a}_{\mathrm{Ir},i}^*$ indicated in Figure~\ref{fig:structure}d are the reciprocal vectors to $\vec{m}_{\mathrm{i}}$, $\vec{a}_{\mathrm{Gr},i}$ and $\vec{a}_{\mathrm{Ir},i}$ (compare Figures~\ref{fig:structure}b and \ref{fig:structure}c). The moiré periodicities derived from LEED are 
$m_1=(31\pm2)$\,\AA\: and $m_2=(9.8\pm0.3)$\,\AA. They agree within the limits of error well with our STM analysis. The same holds for the real space Gr lattice parameter $a_{\mathrm{Gr}}=(2.49\pm0.04)$\,\AA\: derived from the LEED pattern.

On a side note, we also observed the formation of a two-domain Gr phase, but at a lower growth temperature of $1300$\,K. Compare Figure~S3 of the SI for details.

\begin{figure}[!ht]
\centering
\includegraphics[width=8.5cm]{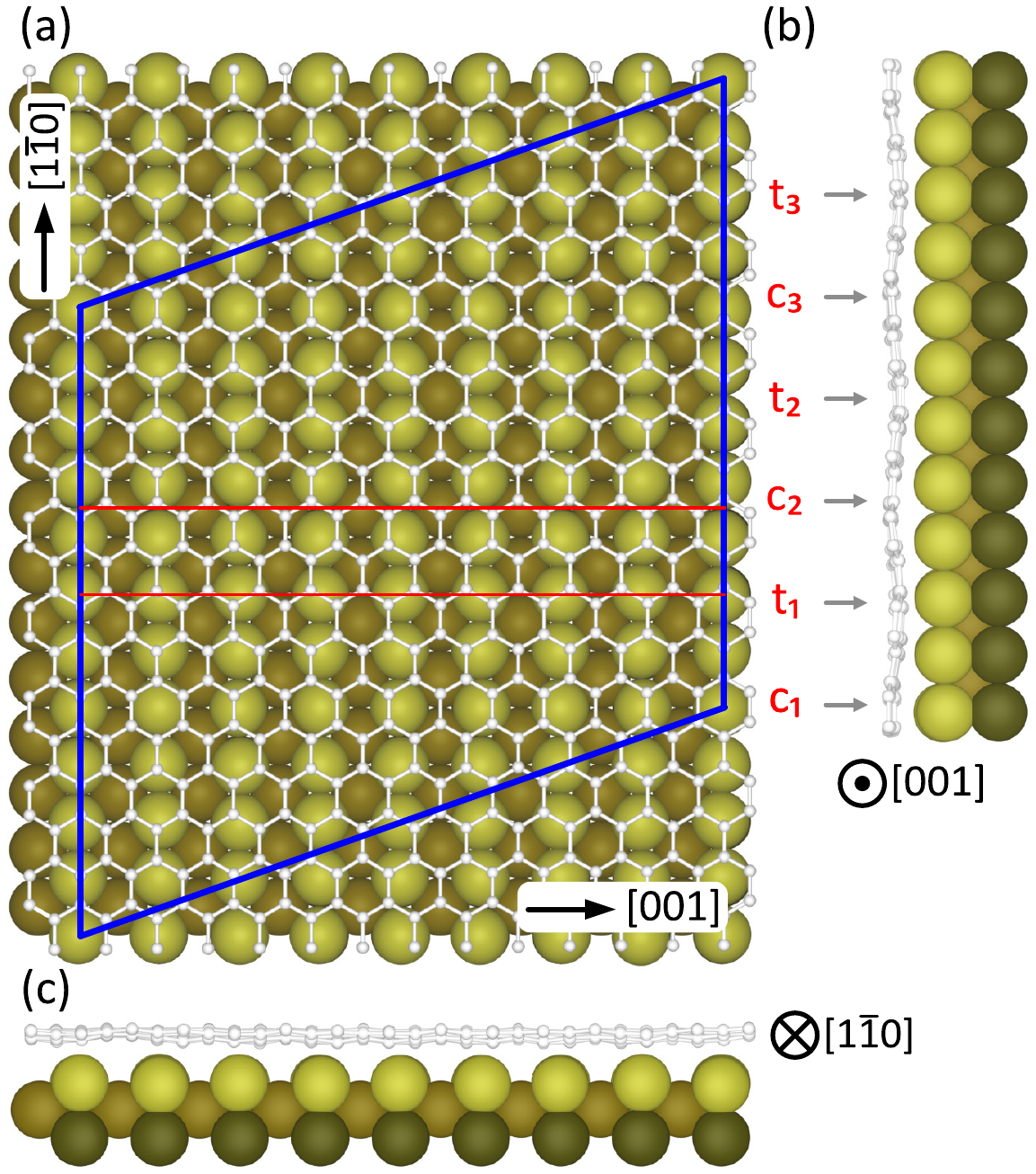}
\caption{Ball model representations of relaxed DFT geometries for Gr/Ir(110). Ir atoms: light (top layer) to dark (bottom layer) brown spheres. C atoms: small light grey dots connected by light grey lines. (a) Top view with DFT supercell indicated by blue rhomboid. The $[1\bar{1}0]$  and $[001]$ directions are also shown. (b) Side view with [001] direction out of drawing plane as indicated. A pattern of crests c$_1$-c$_3$ and troughs t$_1$-t$_3$ in the Gr layer is visible along $[1\bar{1}0]$ with their positions highlighted by arrows. (c) Side view with $[1\bar{1}0]$ direction into drawing plane as indicated.}
\label{fig:DFT_structure}
\end{figure}

Summarizing our analysis, STM and LEED suggest an incommensurate moiré of unstrained or marginally strained Gr with Ir(110). In fact, assuming unstrained Gr with its graphite lattice parameter $a_{\mathrm{graphite}}=2.4612$\,\AA\: and the zigzag rows parallel to the $[001]$ direction, as observed in STM and LEED,  would result in $m_1 = 32.02$\,\AA\: and $m_2=9.91$\,\AA. These numbers agree very well with our STM and LEED analysis. In order to be able to conduct DFT calculations, we need to approximate the experimental situation through a commensurate superstructure cell. The smallest commensurate unit cell with negligible Gr strain is indicated as blue rhomboid in Figure~\ref{fig:structure}c and in the ball model of Figure~\ref{fig:DFT_structure}a. The unit cell is spanned by the primitive translations $\vec{a}_1$ and $\vec{a}_2$ of lengths $a_1=32.58$\,\AA\: and $a_2=29.87$\,\AA, corresponding to 12.5 Gr units on 8 Ir units in $[001]$ direction and 14~Gr rows on 11~Ir units along the $[1\bar{1}0]$ direction. The triple length of $\vec{a}_2$ compared to $\vec{m}_2$ takes into account, that along the $[1\bar{1}0]$ direction approximate commensurability is only achieved after three moiré periods $m_2$. In matrix notation the commensurate superstructure can be expressed as $\bigl(\begin{smallmatrix} 8&4 \\ 0&11 \end{smallmatrix} \bigr)$ with respect to Ir(110) and $\bigl(\begin{smallmatrix} 10&5 \\ -7&14 \end{smallmatrix} \bigr)$ with respect to Gr (compared Figure~S4 in the SI). The Gr lattice parameters $a_{\mathrm{Gr},1}=2.457$\,\AA\: and $a_{\mathrm{Gr},2}=2.463$\,\AA\: in the commensurate superstructure unit cell are close to the in-plane lattice parameters of relaxed graphite $a_{\mathrm{graphite}}=2.4612$\,\AA~\cite{Eckerlin1971} and to our measurements. They are only slightly compressed by $0.17$\,\% or stretched by $0.07$\,\% with respect $a_{\mathrm{graphite}}$.

\textbf{Superstructure -- \textit{ab initio} calculations:} For our DFT calculations we used a slab consisting of 3 layers of Ir, the Gr layer, and 21\,\AA~of vacuum in z direction. The DFT supercell shown in Figure~\ref{fig:DFT_structure}a is based on the superstructure unit cell defined by $\vec{a}_1$ and $\vec{a}_2$ (compare Figure~\ref{fig:structure}c).

The adsorption energy per C atom amounts to $E_{\mathrm{ads}} = -140.2$\,meV, about twice the value obtained for Gr/Ir(111) with the same exchange-correlation functional \cite{FarwickzumHagen16}. Side views of the supercell are presented as Figure~\ref{fig:DFT_structure}b with the direction of view along $[001]$, normal to the dense-packed Ir rows and along a Gr zigzag direction, and Figure~\ref{fig:DFT_structure}c with the direction of view along $[1\bar{1}0]$, i.e. along the dense-packed Ir rows and a Gr armchair direction. The view along $[001]$ displays a clear wave pattern of the Gr layer with wave vector along the direction of $\vec{a}_2$ or the $[1\bar{1}0]$ direction. The wave crests are labeled c$_1$, c$_2$, c$_3$ and the wave troughs are labeled t$_1$, t$_2$, t$_3$. Close inspection reveals that the three wave crests and the three wave troughs are not equivalent in symmetry: while a zigzag row is either precisely aligned to an Ir-atom row along $[001]$ for t$_1$ or between two Ir-atom rows for c$_2$, this alignment is only approximate for the other two crests and troughs. Consequently, the corrugation of the wave pattern is non-uniform ranging from $0.32$\,\AA\: to $0.46$\,\AA. The view along $[1\bar{1}0]$ displays no clear corrugation pattern of the Gr layer. 

These DFT results are in good agreement with the experimental observation of a pronounced wave pattern with crests along $[001]$ with the same periodicity as in DFT. The experimental corrugation of this wave pattern (blue in the height profile of Figure~\ref{fig:structure}b) is generally larger than the corrugation in the direction normal to it (cyan in the height profile of 
Figure~\ref{fig:structure}b), again in qualitative agreement with our DFT calculation. The DFT calculated corrugation of up to $0.46$\,\AA\: is well within the range of experimentally measured corrugations of $0.15$\,\AA\: to $0.8$\,\AA\: along the $[1\bar{1}0]$ direction.

\begin{figure}[!ht]
\centering
\includegraphics[width=12cm]{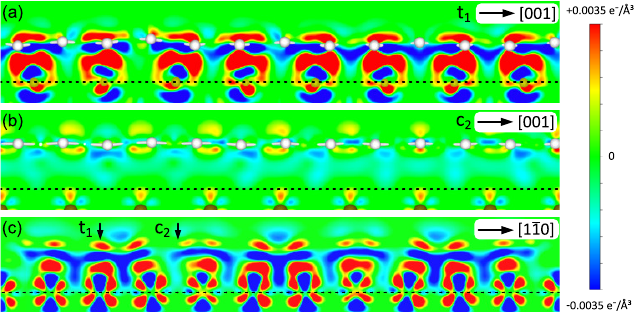}
\caption{(a), (b) Charge-density difference plots in the (a) trough and (b) crest locations indicated in Figure~\ref{fig:DFT_structure}b as t$_1$ and c$_2$ and in Figure~\ref{fig:DFT_structure}a by horizontal red lines.
(c) Charge-density difference average over the entire superstructure unit cell and projected onto a plane along the $[1\bar{1}0]$ direction corresponding to the view of Figure~\ref{fig:DFT_structure}b. The positions of the cuts along t$_1$ and c$_2$ are indicated by arrows. The black dotted lines indicate the position of the top level Ir atoms. See text. Color scale for all plots ranges from charge accumulation in red ($+0.0035$ electrons/\,\AA$^3$) to depletion in blue ($-0.0035$ electrons/\,\AA$^3$).
}
\label{fig:DFT_charge}
\end{figure}

\textbf{Binding -- \textit{ab initio} calculations:} Already the side views of Figure~\ref{fig:DFT_structure}b and \ref{fig:DFT_structure}c suggest that the binding configurations of the C atoms to the Ir(110) substrate vary substantially. On a global level this is evident by noting that the C-Ir bond length, i.e. the distance between a C atom of Gr and the nearest Ir substrate atom, varies between $2.11$\,\AA\: and $3.10$\,\AA, i.e. from a strong chemisorption to a weak chemisorption bond length. This large variation is primarily caused by the large corrugation of the Ir substrate with its hill and valley structure. A manifestation of the binding heterogeneity is the Gr wave pattern as apparent in the side view of Figure~\ref{fig:DFT_structure}b. To obtain insight into the underlying physics and associated wavelike variation of Gr's properties, charge-density difference plots along the red lines in Figure~\ref{fig:DFT_structure}a in the trough t$_1$ and the crest c$_2$ are compared in Figures~\ref{fig:DFT_charge}a and \ref{fig:DFT_charge}b, respectively. In the cut through the trough t$_1$ shown in Figure~\ref{fig:DFT_charge}a charge accumulation (red) between the Ir and the C atoms signals the formation of chemical bonds. These chemical bonds to the substrate cause the C atoms of Gr to acquire partial sp$^3$ character. Moreover, charge accumulates in the $\pi$ system above the C atom plane. In the cut through the crest c$_2$ shown in Figure~\ref{fig:DFT_charge}b essentially no charge accumulation between the C atoms and the distant Ir atoms is present -- chemical bonds are weak. Therefore, contrary to the troughs, in the crests little charge is injected into the $\pi$ system above the C atoms. Besides a variation in local work function, the variation of the local charge transfer into the Gr $\pi$ system implies also a variation of the van der Waals interactions of the Gr layer with physisorbed species. As outlined in~\cite{Huttmann2015}, the strength of the van der Waals interactions is larger where the charge cloud of the $\pi$ system spreads out into the vacuum further away from the C nuclei. The modulation of the binding character and electronic properties with the wave pattern is also apparent in Figure~\ref{fig:DFT_charge}c, which shows the charge-density difference average over the entire superstructure unit cell and projected onto a plane along the $[1\bar{1}0]$ direction. There, the variation of binding character and charge donation to the Gr $\pi$ system is well visible through the variation of strong charge accumulation above and below the C atom plane.  

The charge density difference plots normal to the wave pattern along the Gr armchair direction ($[1\bar{1}0]$ direction of the Ir substrate) also display a variation of electronic properties. However, on all cuts the dominant wave pattern is superimposed, causing a considerable heterogeneity along the direction of cut. Compare also Figure~S5 of the SI.  

\begin{figure}[!ht]
\centering
\includegraphics[width=12cm]{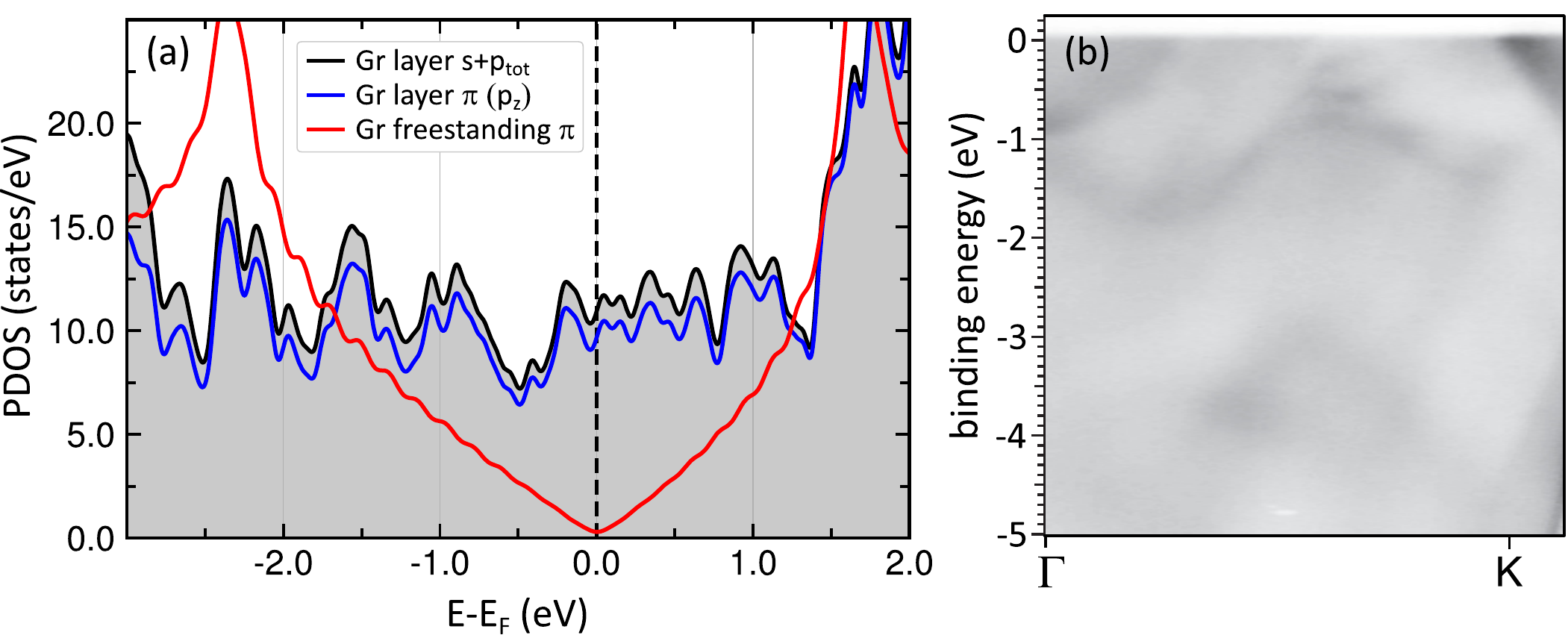}
\caption{(a) Red line: Density of states (DOS) of freestanding Gr. Black line and gray-shaded area: Gr partial DOS when on Ir(110). Blue line: Gr partial DOS projected on the Gr $\pi$ system consisting of p$_\mathrm{z}$ atomic-like orbitals. See text. (b) Angle-resolved photo emission spectrum along $\Gamma-\mathrm{K}$, recorded at a photon energy of $100$\,eV and a temperature of $T=35$\,K. There is no Dirac cone at the K point and there are no features that could be related to the Gr $\pi$ bands in agreement with the DFT calculations.}
\label{fig:ARPES_DOS}
\end{figure}

\textbf{Electronic structure of Gr on Ir(110):} Figure~\ref{fig:ARPES_DOS}a compares the characteristic V-shaped freestanding Gr density of states (red line) with the Gr partial DOS when adsorbed to Ir(110) (black line). The V-shape of the freestanding Gr DOS signals electronically intact Gr with a Dirac cone formed by the Gr $\pi$ and $\pi^*$ bands that touch at the Dirac point, where the DOS vanishes. This feature is characteristic for freestanding or physisorbed Gr layers and corresponds to an sp$^2$ hybridization of its electronic states. The absence of this feature and the substantial partial Gr DOS of Gr/Ir(110) in the entire energy range of a few eV around the Fermi energy signifies considerable modification of the Gr electronic structure when adsorbed to Ir(110), similar e.g. to the case of Gr on Ni(111) \cite{Wang13}. The diversity of C-Ir bonds noticed already in Figure~\ref{fig:DFT_structure} gives rise to a diversity of C-Ir hybridizations, which together with the variation of the local charge transfer visible in Figure~\ref{fig:DFT_charge}, gives rise to the smeared out partial DOS of Gr. The blue curve in Figure~\ref{fig:ARPES_DOS}a represents the partial Gr DOS projected onto the carbon p$_\mathrm{z}$ atomic-like orbitals that originally form the Gr $\pi$ system. The difference between this projection and the entire partial Gr DOS is small, but indicates a non-negligible sp$^3$ character of bonding due to the local chemical interactions between the corrugated Gr and Ir(110) as depicted in Figure~\ref{fig:DFT_charge}.

The detailed electronic structure of the adsorption system has been determined by ARPES~\cite{Hoffmann2004}. Figure~\ref{fig:ARPES_DOS}b shows the photoemission intensity as a function of binding energy and $k_{\parallel}$ in the $\Gamma-\text{K}$ direction of the Gr Brillouin zone. The ARPES data show no sign of a Dirac cone nor any feature that could be related to the Gr $\pi$ band, fully consistent with the DFT calculations. See Figure~S6 in the SI and the related discussion for a more detailed analysis.

\textbf{A naphthalene molecule as sensor for the energy landscape of adsorption on Gr/Ir(110):} To obtain insight to whether the modulation associated to the wave pattern of Gr/Ir(110) can be used to template molecular adsorption, a naphthalene molecule was employed as sensor in DFT calculations of still feasible computational effort. Using different starting configurations for C$_{10}$H$_{8}$ adsorbed to the crest c$_2$ and trough t$_1$ of Gr/Ir(110) (compare Figure~\ref{fig:DFT_structure}) several local minima of adsorption energy were identified.

\begin{figure}[!ht]
\centering
\includegraphics[width=16cm]{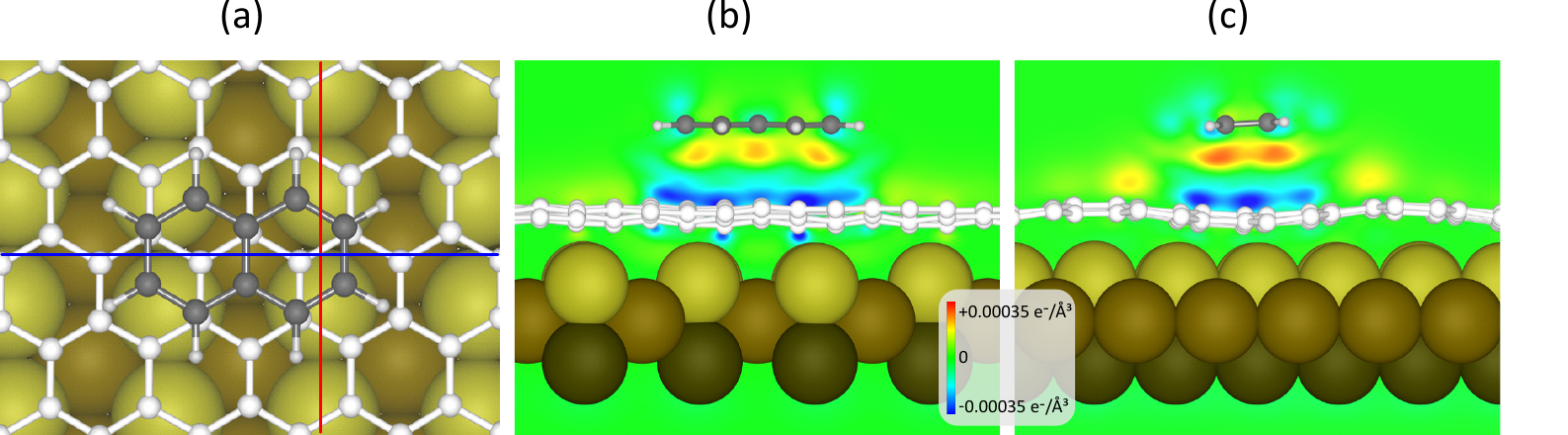}
\caption{(a) Top view ball model for minimum energy adsorption geometry of C$_{10}$H$_{8}$ in trough t$_1$ on Gr/Ir(110) (compare Figure~\ref{fig:DFT_structure}). (b) Charge-density difference plot along blue line in (a) of combined system compared to sum of isolated C$_{10}$H$_{8}$ molecule and Gr/Ir(110). (c) Same as (b), but along red line in (a). Color scale in (b) and (c) is the same as in Figure~\ref{fig:DFT_charge}, but used for one order of magnitude smaller charge density-differences ranging from $+0.00035$ electrons/\,\AA$^3$ for charge accumulation in red to $-0.00035$ electrons/\,\AA$^3$ for charge depletion in blue.}
\label{fig:DFT_Nph}
\end{figure}

Figure~\ref{fig:DFT_Nph}a displays the minimum energy adsorption geometry, where the molecule resides in the trough, as also obvious from the side view cuts of the charge-density difference plots of Figure~\ref{fig:DFT_Nph}b and \ref{fig:DFT_Nph}c. Adsorption takes place with the long molecular axis along the trough. No energy minimum could be found for the molecule in different orientations, e. g. with the long axis normal to the trough. The adsorption energy $E_{\mathrm{ads}}$ is -947\,meV, lower by 168\,meV compared to the best-bound configuration on a crest. The minimum energy adsorption site in the trough corresponds to the locations where the charge accumulation above Gr is highest (compare Figure~\ref{fig:DFT_charge}). Thus these results are in qualitative agreement with experiments and DFT calculations for naphthalene adsorbed to Gr/Ir(111) \cite{Huttmann2015}, where the strength of the van der Waals interactions was found to increase with the charge donated to Gr (n-doping). For the minimum energy configuration shown in Figure~\ref{fig:DFT_Nph}, the average naphthalene-graphene distance is $3.27$\,\AA, a distance typical for a physisorbed molecule, while for all other local adsorption energy minima the distances are larger. Compare 
Figure~S7 and Table~S1 of the SI for additional calculations and more details.

Our calculations make plain that despite Gr's strong interaction with the Ir(110) substrate, it is still an inert substrate for molecular adsorption, acting as a spacer effectively separating the metal from the molecule. The situation is not unexpected, when considering that also other `strongly' interacting Gr layers act still as inert spacer between molecules and the underlying metal. To give an example, the peak desorption temperature of benzene from `strongly' interacting Gr on Ru(0001)~\cite{Batzill2012} is within the limits of error identical to the one from graphite~\cite{Chakradhar2013}.

\textbf{Use of Gr/Ir(110) as nanotemplate for alignment in on-surface synthesis:} Our theoretical calculations for C$_{10}$H$_{8}$ adsorbed to Gr/Ir(110) imply that this substrate displays an anisotropic physisorption energy landscape and is thereby able to template adsorption and to impose uniaxial alignment. Here, we use the example of on-surface synthesis of sandwich-molecular wires to demonstrate these properties. A sandwich-molecular wire is an organometallic compound consisting of an alternation of metal atoms with ring-shaped aromatic molecules \cite{Hosaya14,Nakajima00}. When Eu atoms and cyclooctatetreaene (Cot, C$_8$H$_8$) molecules (eight-membered carbon rings) are combined in room temperature on-surface synthesis on Gr/Ir(111), Eu is evaporated onto the substrate in a background pressure of Cot, of which the excess re-evaporates at 300\,K \cite{Huttmann2017}. Because of van der Waals interaction between the wires, they interlock and form monolayer high, \textit{randomly} oriented islands of parallel wires resulting in diffraction rings rather than spots in LEED \cite{Huttmann2017,Huttmann2019}. The lack of island orientation was found not to depend on coverage.

\begin{figure}[!ht]
\centering
\includegraphics[width=16cm]{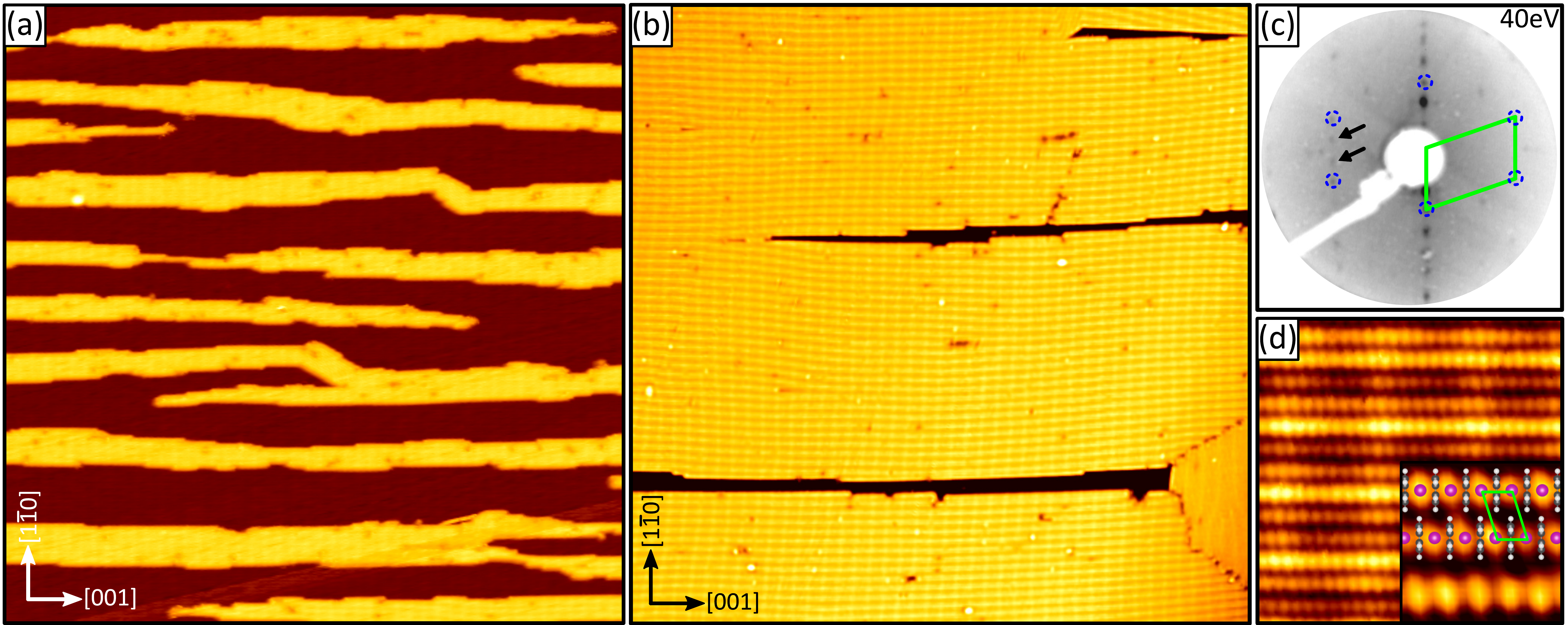}
\caption{(a), (b) STM topographs ($1500 \times 1500$\,\AA$^2$) of (a) elongated EuCot wire carpet islands and (b) of a full layer on Gr/Ir(110) oriented along the $[001]$ direction. (c) $40$\,eV LEED pattern of the same sample as in (b). The unit cell of the EuCot wire carpet is indicated by green lines, with the corresponding first order reflections encircled dashed-blue. $3 \times 1$ superstructure reflections are highlighted by black arrows. Note that different LEED set-ups cause different sizes of LEED patterns. (d) STM topograph ($100 \times 100$\,\AA$^2$) of EuCot/Gr/Ir(110) with a $3 \times 1$ intensity variation along the $[1\bar{1}0]$ direction resulting from the lattice mismatch of the substrate moiré and the wire carpet. Inset: molecular resolution STM topograph ($25 \times 25$\,\AA$^2$) overlayed with a wire model. Magenta dots: Eu atoms. White and black dots: H atoms and C atoms of Cot. The wire carpet unit cell is indicated as green rhomboid. STM topographs taken at $300$\,K. Tunneling parameters are (a) $U_{\mathrm{bias}}=-2.0$\,V and $I_{\mathrm{t}}=0.03$\,nA, (b) $U_{\mathrm{bias}}=-1.74$\,V and $I_{\mathrm{t}}=0.08$\,nA, and (d) $U_{\mathrm{bias}}=-1.82$\,V and $I_{\mathrm{t}}=0.35$\,nA.}
\label{fig:wires}
\end{figure}

As obvious from Figure~\ref{fig:wires}, successful on-surface synthesis of EuCot is also possible on Gr/Ir(110). This observation implies directly the mobility of reaction intermediates to wire ends and the re-evaporation of the Cot excess at room temperature.
This underpins the inertness of the Gr/Ir(110) substrate and the physisorbed state of the unreacted molecules. In contrast to the random orientation of the wire carpet islands on Gr/Ir(111), growth on the anisotropic Gr/Ir(110) substrate gives rise to thin and long wire islands all oriented along the $[001]$ direction as visible in Figure~\ref{fig:wires}a. For larger coverages a well-oriented, coalesced monolayer results, as shown in Figure~\ref{fig:wires}b. Only step edges along $\left[1\bar{1}0\right]$ may cause a deviation from the global $[001]$ alignment in small patches, e.g. in the lower right of Figure~\ref{fig:wires}b. Figure~\ref{fig:wires}c displays the corresponding LEED pattern. It exhibits clear diffraction spots of the wire lattice encircled in dotted-blue. The rhomboidal reciprocal unit cell is indicated in green. Because LEED is a spatially averaging technique, the diffraction pattern implies a global alignment of the wires along the $[001]$ direction. Highlighted by the two black arrows in Figure~\ref{fig:wires}c, $3 \times 1$ superstructure reflections can be identified in the direction normal to the wires, of which the origin is explained below. In the molecular resolution STM topograph of Figure~\ref{fig:wires}d the parallel wires are shown together with a ball model overlay of the molecular structure and the wire lattice unit cell. Close inspection of the STM topograph reveals a beating of the wire height (brightness) along $[1\bar{1}0]$, where about every third wire appears to be higher. It is this height variation of the wires that gives rise to the $3 \times 1$ superstructure spots in LEED, which is thus a global feature of wire ordering. The intensity variation can be explained by the mismatch of the interwire distance and the moiré periodicity of the underlying substrate in this direction. Based on the crest spacing of $\approx 10$\,\AA\: and the wire spacing of $\approx 6.8$\,\AA\: measured on Gr/Ir(111) \cite{Huttmann2017}, three wires fit on two moiré periodicities along $\vec{m}_2$. Apparently two thirds of the wires are located close to the trough positions, while one third is located close to a crest position. Furthermore, faint vertical lines of brighter contrast spaced by $m_1$ reflect the bending of the horizontally aligned wires over the moiré periodicity along $[001]$.

Based on the successful on-surface synthesis using the same parameters as for EuCot wire growth on Gr/Ir(111) \cite{Huttmann2017,Huttmann2019}, it is evident that also on Gr/Ir(110) the wires are bound through van der Waals interactions to the substrate. The upright standing aromatic Cot-dianions are in contact with Gr only through their peripheral H-atoms. Bound to the cyclic carbon ring, they are unable to interact chemically with Gr. For a single wire, our DFT calculation shown above suggests adsorption to a trough location, where the van der Waals interaction is stronger than on the crests.
However, due to substantial interwire van der Waals interaction, single wires are not realized even for smaller coverages. Nevertheless, consistent with a preferential binding to the troughs, two thirds of the wires are adsorbed close to the throughs rather than to the crests, as noticed above when discussing the $3 \times 1$ wire superstructure.   

Elastic energy considerations are also in favor of adsorption along the wave pattern, i.e. the $[001]$ direction, rather than vertically to the wave pattern. In order to maximize the binding to the substrate, the wires need to adhere conformal to the Gr-sheet in an optimum distance defined by Pauli repulsion and van der Waals interactions. For a 1D-wire oriented perpendicular to the wave pattern along the $[1\bar{1}0]$ direction this would imply substantially more bending with shorter periodicity to conform to the wave pattern than for a wire oriented along the $[001]$-direction with substantially less corrugation and larger periodicity $m_1$. Thus, the elastic energy penalty for wire orientation along the $[001]$ direction is lower, also favoring its orientation along this direction.

It is remarkable that physisorbed species -- the 1D sandwich-molecular wires -- are perfectly oriented through the Gr/Ir(110) template at temperatures as high as room temperature. However, even if the charge modulation and elastic energy effects are presumably small per formula unit -- possibly as low as $10$\,meV -- the wires are composed of hundreds of formula units. Thereby, energy differences for wires adsorbed in different orientations and at different adsorption sites become large.

\begin{figure}[!ht]
\centering
\includegraphics[width=10cm]{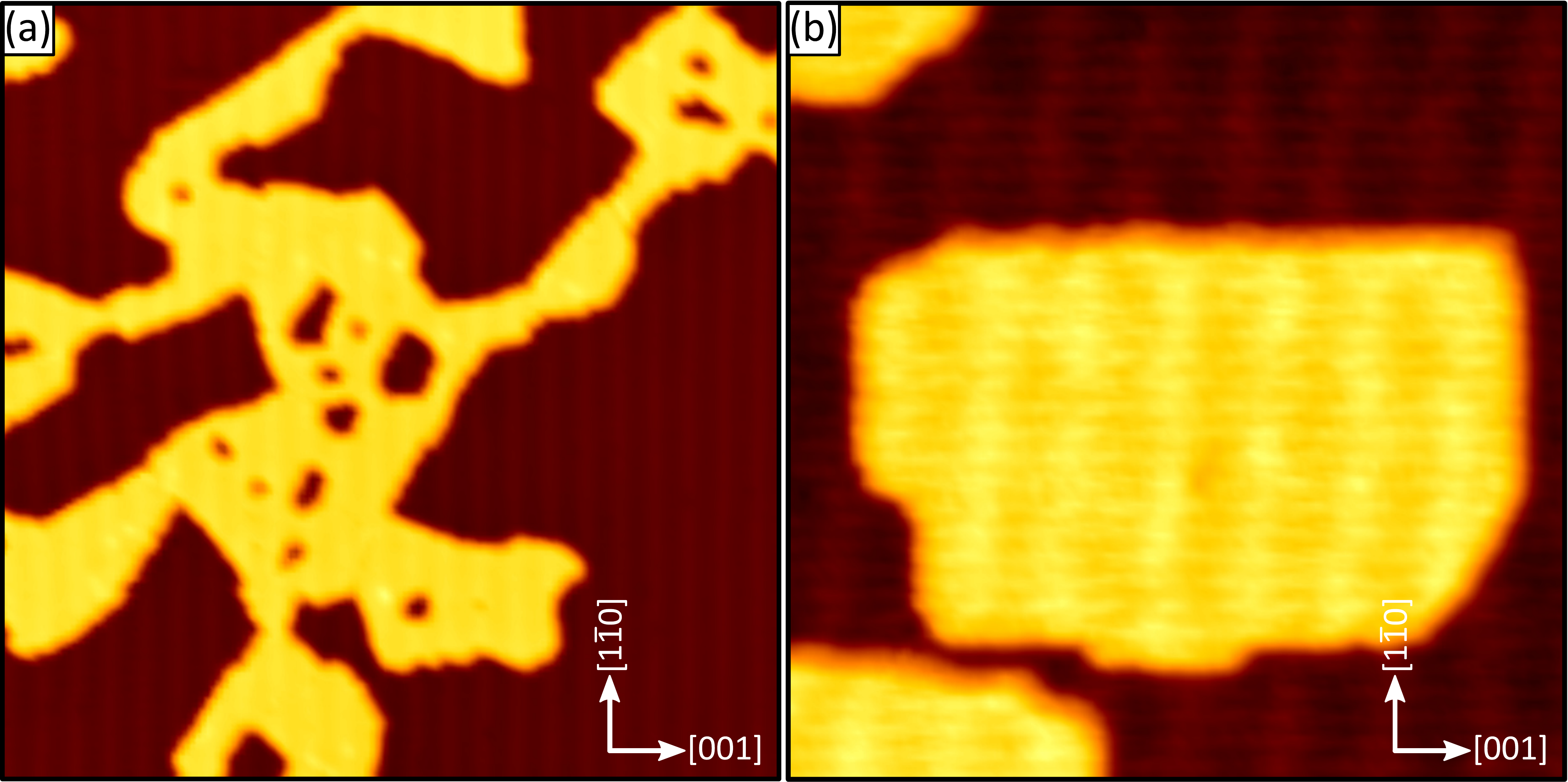}
\caption{(a) STM topograph ($750 \times 750$,\AA$^2$) taken after MBE growth of monolayer NbS$_2$ islands on Gr/Ir(110) at 150\,K and additional annealing to 825\,K. (b) STM topograph ($300 \times 300$\,\AA$^2$) of a pseudomorphic Fe intercalation island underneath Gr on unreconstructed Ir(110). STM topographs taken at $1.7$\,K. Tunneling parameters are (a) $U_{\mathrm{bias}}=2.0$\,V and $I_{\mathrm{t}}=0.1$\,nA and (b) $U_{\mathrm{bias}}=1.0$\,V and $I_{\mathrm{t}}=0.5$\,nA.}
\label{fig_application}
\end{figure}

\textbf{Additional uses of Gr/Ir(110) as substrate:} Figure~\ref{fig_application}a displays monolayer NbS$_2$ islands of excellent structural quality grown on Gr/Ir(110) through MBE following the method described in \cite{Hall2019}. Besides Gr/SiC(0001) \cite{Ugeda2014} and Gr/Ir(111) \cite{Hall2018}, Gr/Ir(110) is thus a third inert Gr system suitable for the MBE growth of 2D layers and 2D layer heterostructures. The orientation of the transition metal dichalcogenide (TMD) islands is close to random after room temperature growth and mild annealing, not different to TMD growth on the other Gr substrates. The random orientation indicates a weak interaction of the TMD with Gr/Ir(110).

Figure~\ref{fig_application}b displays Fe islands intercalated underneath Gr on Ir(110). The unique feature here is that the Gr cover enables epitaxial growth on unreconstructed Ir(110) which has not been possible before. A pseudomorphic Fe monolayer on Ir(110) as realized here could offer magnetic properties similarly exciting as those of the Fe monolayer on Ir(111)~\cite{Heinze2011}. 

\textbf{Discussion:}
Instead of forming the nano-facet reconstruction, Ir(110) remains unreconstructed upon cooldown to room temperature when Gr has been grown on it at 1500\,K. To explain this remarkable observation we note that the surface reconstructions of Ir(110) are driven by the temperature-dependent minimization of surface free energy $\gamma$ which itself is linked to the surface stress \cite{Shuttleworth1950}. The Ir(110) surface was shown to run through a sequence of reconstructions upon cooling with a $(2 \times 1)$ missing row reconstruction being present at $800$\,K--$900$\,K, a $(3 \times 1)$ missing row reconstruction being dominant at $500$\,K--$600$\,K, while eventually upon cooling the $(331)/(33\bar{1})$ nano-facet reconstruction forms~\cite{Ney02,Schulz2000}. The surface structure of Ir(110) at the Gr growth temperature of $1500$\,K is unknown, but based on the results for the same surface orientation of the parent element Pt~\cite{Robinson1990}, at $1500$\,K Ir(110) is presumably above the roughening temperature and will display fast surface profile fluctuations due to high mobility. Knowing that a Gr membrane replicates the surface morphology at the growth temperature \cite{Kraus13}, the morphology observed at $300$\,K by STM is close to the morphology at the end of growth at $1500$\,K. Therefore, we tentatively conclude that Gr grows on unreconstructed Ir(110). The calculated binding energy of $140.2$\,meV per C atom is of similar magnitude as the Ir surface free energy, which can be estimated to be of the order of $300$\,meV for Ir(110) if normalized to the Gr atomic density \cite{Tran2016}. It is therefore plausible that upon cooling, reconstructions that lower the surface energy in the clean case, are suppressed as they would diminish the adhesion between Gr and the Ir(110) substrate. In fact, any of the known Ir(110) reconstructions make the surface rougher and thus either would reduce the number of binding substrate atoms or force large deformations in Gr to conform to the substrate. In brief and somewhat coarse, the adhering Gr layer shifts the binding of Ir(110) surface atoms to a more bulk-like situation due to the formation of Ir-C bonds, largely relieving the driving force for surface reconstruction. 

Gr on Ir(110) is a single crystal, if grown at $1500$\,K, while two domains are found for growth at $1300$\,K (compare Figure~S3 of the SI). The situation is similar for the growth of Gr on Ir(111), where upon increasing the growth temperature the situation changes from multiple domain orientations to a single crystal of Gr \cite{Loginova09,Hattab11}. The lack of strain in the $\bigl(\begin{smallmatrix} 8&4 \\ 0&11 \end{smallmatrix} \bigr)$ commensurate approximation of the high-temperature superstructure suggests that in fact the low-temperature two-domain structure which is considerably strained (compare Figure~S3 of the SI and related discussion) reflects kinetic limitations of the Gr growth process rather than a change in the orientation-dependent adsorption energy. To substantiate such speculations systematic growth temperature dependent studies would be necessary, which are beyond the scope of this work. Nevertheless, we speculate that if the growth temperature can be raised sufficiently, also on other fcc(110) surfaces single-crystal Gr layers could be grown.

Lastly, one might ask why Gr on Ir(110) is so different from Gr on Ir(111), with a much higher binding energy and an electronic structure lacking a Dirac cone. As a first remark it should be noted that although the binding energy of graphene to Ir(110) is with $140.2$\,meV per C-atom much larger than the $69$\,meV per C-atom to Ir(111) \cite{FarwickzumHagen16}, it is still much lower than typical chemisorption energies of several eV per molecule or the $7.6$\,eV cohesive energy of Gr \cite{LandoltB}. The binding energy of Gr to Ir(110) is comparable to the binding energy of Gr to Ni(111) \cite{Silvestrelli2015}, which is considered as weak chemisorption \cite{Mittendorfer2011}. We also note, that the electronic structure of Gr is especially sensitive to the environment because of the low density of states close to the Dirac point. Therefore a loss of the Dirac cone, as also observed for Ni(111) \cite{Wang13} does not imply a loss of the predominat sp$^2$ bonding character, as obvious from the projected DOS in Figure~\ref{fig:ARPES_DOS}.

Compared to Ir(111), which is smooth and where all surface atoms are 9-fold coordinated, Ir(110) is a more open and a corrugated surface, with the surface atoms in the protruding rows being only 6-fold coordinated, whereas the surface atoms in the troughs are 1.36\,\AA~below the level of the row atoms and 11-fold coordinated. Evidently, on Ir(110) the surface atoms with lower coordination are more reactive. For instance, DFT calculations show that the CO binding energy on Ir(110) is larger by 410\,meV compared to Ir(111) in the low coverage limit \cite{Liu2019}, a difference much larger than the 71\,meV difference in binding energy of Gr to Ir(110) and Ir(111).

We speculate that the overall van der Waals interaction pulls Gr towards the surface such that the protruding atoms on Ir(110) start to hybridize with the Gr layer, while the recessed atoms do not. On the flat Ir(111) surface, Pauli repulsion stops the approach prior to the onset of significant hybridization of specific substrate orbitals with the Gr sheet \cite{Busse2011}. Based on this proposed scenario, we speculate that a similar difference in binding also holds for Gr on other metal surfaces, e.g. for Pt.

\section{Conclusion}

Single-domain Gr on Ir(110) forms upon low-pressure CVD growth at $1500$\,K, displays a rectangular moiré pattern with periodicities $m_1 = 33$\,\AA\: in $[001]$ direction and $m_2 = 10$\,\AA\: in $[1\bar{1}0]$ direction, and can be approximated as a $\bigl(\begin{smallmatrix} 8&4 \\ 0&11 \end{smallmatrix} \bigr)$ superstructure with respect to Ir(110). The Gr layer is chemisorbed to Ir(110) with an adsorption energy of $-140.2$\,meV per C atom. Due to strong and locally varying interaction with the substrate it lacks a Dirac cone.

The Gr layer displays a wave pattern with wave vector in $[001]$ direction and corrugation of 
$\approx 0.4$\,\AA\: according to our \textit{ab initio} calculations. This wave pattern implies a modulation in charge transfer to the Gr $\pi$ system and in the Gr-Ir hybridization, both being most pronounced in the trough of the wave pattern. The effect of this property modulation on the physisorption of aromatic molecules is explored through \textit{ab initio} calculations for a naphthalene molecule. The adsorption energy landscape is found to be highly anisotropic with the maximum binding energy for molecules with their long axis adsorbed along the troughs, where the van der Waals interaction is strongest because of the larger transferred charge to Gr. The same property modulation is shown experimentally to enable the alignment of EuCot sandwich-molecular wires along troughs at $300$\,K. The successful on-surface synthesis -- requiring re-evaporation of excess Cot molecules and the diffusion of reaction intermediates -- also documents the inertness of the substrate. This property is also at the heart of the successful use of Gr/Ir(110) as a substrate for the growth of the quasi-freestanding transition metal dichalcogenide layer NbS$_2$ through reactive MBE. 

Under the Gr cover, Ir(110) remains unreconstructed down to the lowest temperatures. It is argued that the strong adhesion between Gr and Ir(110) suppresses the formation of the nano facets, $(2 \times 1)$, and $(3 \times 1)$ reconstructions. By intercalation thereby epitaxial layers can be grown on unreconstructed Ir(110) as exemplified for a pseudomorphic Fe monolayer. 

\section*{Acknowledgements}

This work was funded by the Deutsche Forschungsgemeinschaft (DFG, German Research Foundation) within the project 'Sandwich molecular nanowires: on-surface synthesis, structure and magnetism' (MI 581/23-1, AT 109/5-1 and WE 2623/17-1). V. C., N. A., J. F., T. K., and T. M acknowledge additional DFG support within CRC1238, project no. 277146874 - CRC 1238 (subprojects C01 and B06). We gratefully acknowledge the Gauss Centre for Supercomputing (GCS) for providing computing time through the John von Neumann Institute for Computing (NIC) on the GCS share of the supercomputer JURECA at J\"ulich Supercomputing Centre (JSC). This work was also supported by VILLUM FONDEN via the Centre of Excellence for Dirac Materials (Grant No. 11744).

\section*{Author contributions}
S.K, F.H., J.F., T.K. and K.B. conducted the syntheses and the STM and LEED experiments. S.K., M.B., R.-M.S., A.J.H. and A.H. conducted the ARPES experiments. N.A., S.T. and V.C. conducted the theoretical calculations. S.K., N.A. and T.M. wrote the manuscript with contributions from all authors. All authors contributed to the scientific discussion.

\section*{Supporting information available}
The supporting information provides data for pristine reconstructed Ir(110) (Figure~S1), the Fourier transform of Figure~\ref{fig:structure}c (Figure~S2), data and a structure model for the $T=1300$\,K two-domain phase of Gr/Ir(110) (Figure~S3), an illustration (Figure~S4) and explanation for the matrix notation of the approximate commensurate superstructure cell, additional charge-density difference plots (Figure~S5), and additional ARPES data in Figure~S6 with further evidence for the absence of a Gr Dirac cone. Figure S7 displays the geometries of all adsorption energy minima found for C$_{10}$H$_{8}$ on Gr/Ir(110), and Table~S1 the details of adsorption in these minima.

\section*{References}
\bibliography{GrIr110}

\clearpage

\title{Supporting information: Single-crystal graphene on Ir(110)}

\author{Stefan Kraus$^1$, Felix Huttmann$^1$, Jeison Fischer$^1$, Timo Knispel$^1$ , Ken Bischof$^1$, Alexander Herman$^2$, Marco Bianchi$^3$, Raluca-Maria Stan$^3$, Ann Julie Holt$^3$, Vasile Caciuc$^4$, Shigeru Tsukamoto$^4$, Heiko Wende$^2$, Philip Hofmann$^3$, Nicolae Atodiresei$^4$, and Thomas Michely$^1$}
\address{$^1$ II. Physikalisches Institut, Universität zu K\"{o}ln, Zülpicher Str. 77, 50937 K\"{o}ln, Germany}
\address{$^2$ Faculty of Physics and Center for Nanointegration Duisburg-Essen (CENIDE), University of Duisburg-Essen, Lotharstraße 1, 47048 Duisburg, Germany}
\address{$^3$ Department of Physics and Astronomy, Interdisciplinary Nanoscience Center (iNANO), Aarhus University, 8000 Aarhus C, Denmark}
\address{$^4$ Peter Gr{\"{u}}nberg Institute and Institute for Advanced Simulation, Forschungszentrum Jülich, Wilhelm-Johnen-Straße, 52428 J{\"{u}}lich, Germany}
\ead{kraus@ph2.uni-koeln.de (experiment); n.atodiresei@fz-juelich.de (theory)}

\vspace{10pt}
\begin{indented}
\item[]\today
\end{indented}


\maketitle
 
\newpage
\section*{Figure S1: Surface reconstruction of Ir(110)}

\begin{figure}[!h]
\centering
\includegraphics[width=11cm]{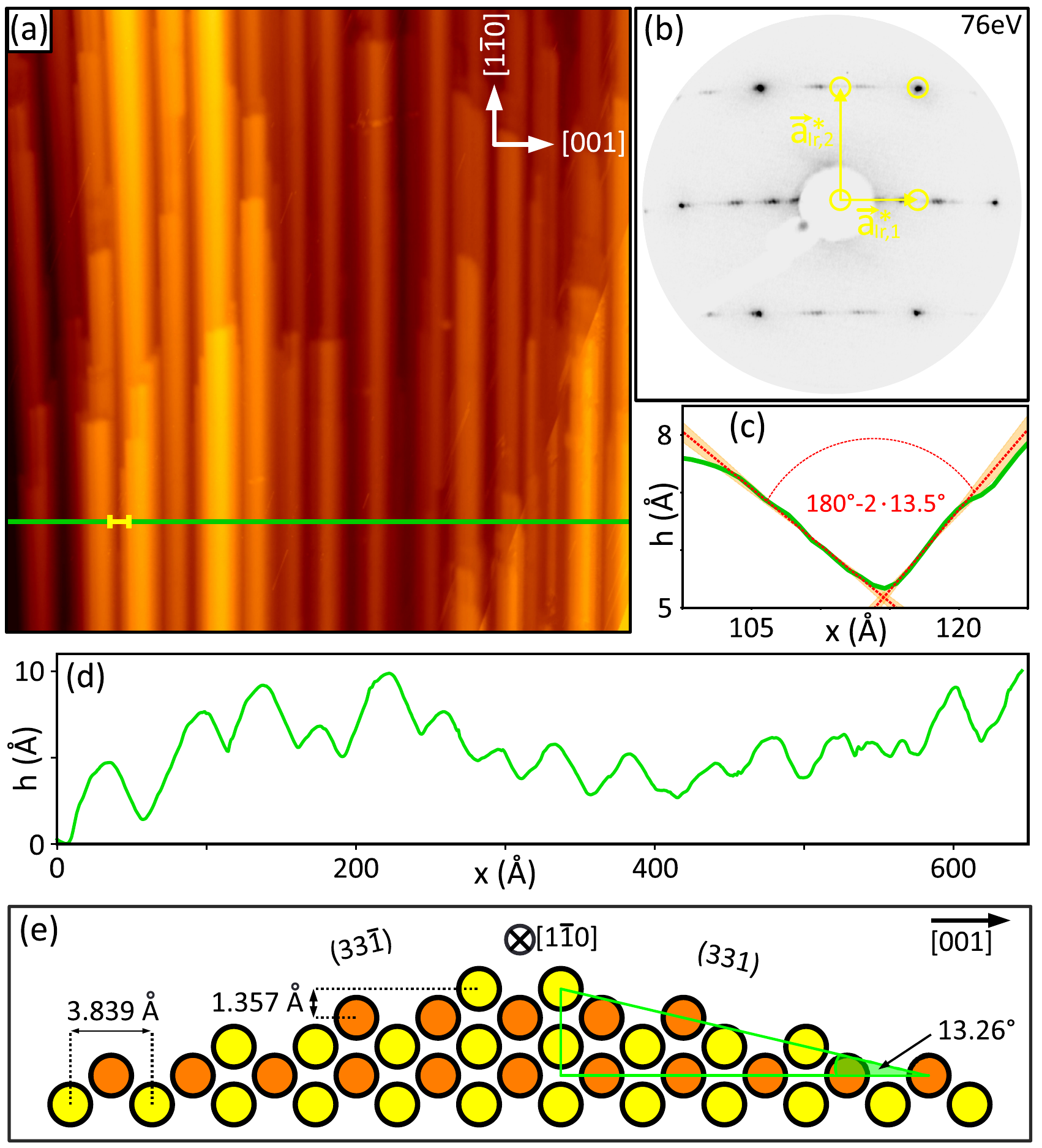}
\caption{(a) STM topograph ($650 \times 650$\,\AA$^2$) of facetted, clean Ir(110) without Gr. The $[001]$ and $[1\bar{1}0]$ directions are specified in the upper right corner and hold also for (b). (b) $76$\,eV contrast-inverted LEED pattern of sample in (a). Locations of Ir reflections for unreconstructed Ir(110) are encircled yellow. First order reflections are split due to facetting. (c) Height profile along yellow line segment in (a). The slopes are indicated by dotted-red lines, the corresponding uncertainties as orange cones. The angle between the slopes is $153$\,° and agrees well with the $(331)$ and $(33\bar{1})$ nano-facets being inclined by $13.26$\,° with respect to the (110) plane. (d) Height profile along green line in (a). Side view ball model of surface displaying $\left(331\right)$ and $(33\bar{1})$ nano-facets consistent with the profiles of (c) and (d). STM imaging temperature in (a) is $300$\,K, tunneling parameters in (a) are $U_{\mathrm{bias}}=-1.80$\,V and $I_{\mathrm{t}}=0.35$\,nA.}
\label{fig:reconstruction}
\end{figure}

We have investigated the reconstructed surface of an iridium (110) single crystal as previously discussed in~[1,2]. Figure~\ref{fig:reconstruction}a shows an STM topograph, in which the reconstruction is visible. The surface is not flat, but forms elongated ridges along the $[1\bar{1}0]$-direction. In the LEED pattern of Figure~\ref{fig:reconstruction}b the Ir(110) reflections and reciprocal lattice translations are indicated in red. The first order reflections are split into pairs of two spots centered around the regular positions along the $[001]$ direction, while the second order reflexes are not split. Figure~\ref{fig:reconstruction}c shows the contact point of two ridges as indicated by the yellow line in \ref{fig:reconstruction}a. The measured contact angle is $(13.5\pm0.5)$\,°. The profile in Figure~\ref{fig:reconstruction}d shows height variations between the ridges of less than $10$\,\AA\: and distinct contact angles between these ridges. The formation of $(331)$ and $(33\bar{1})$ facets reduces the surface energy and leads to an angle of $13.26$\,° with respect to the (110) surface~[1].

\newpage
\section*{Figure S2: Fourier transform of atomic resolution STM topograph of Gr/Ir(110)}

\begin{figure}[!ht]
\centering
\includegraphics[width=10cm]{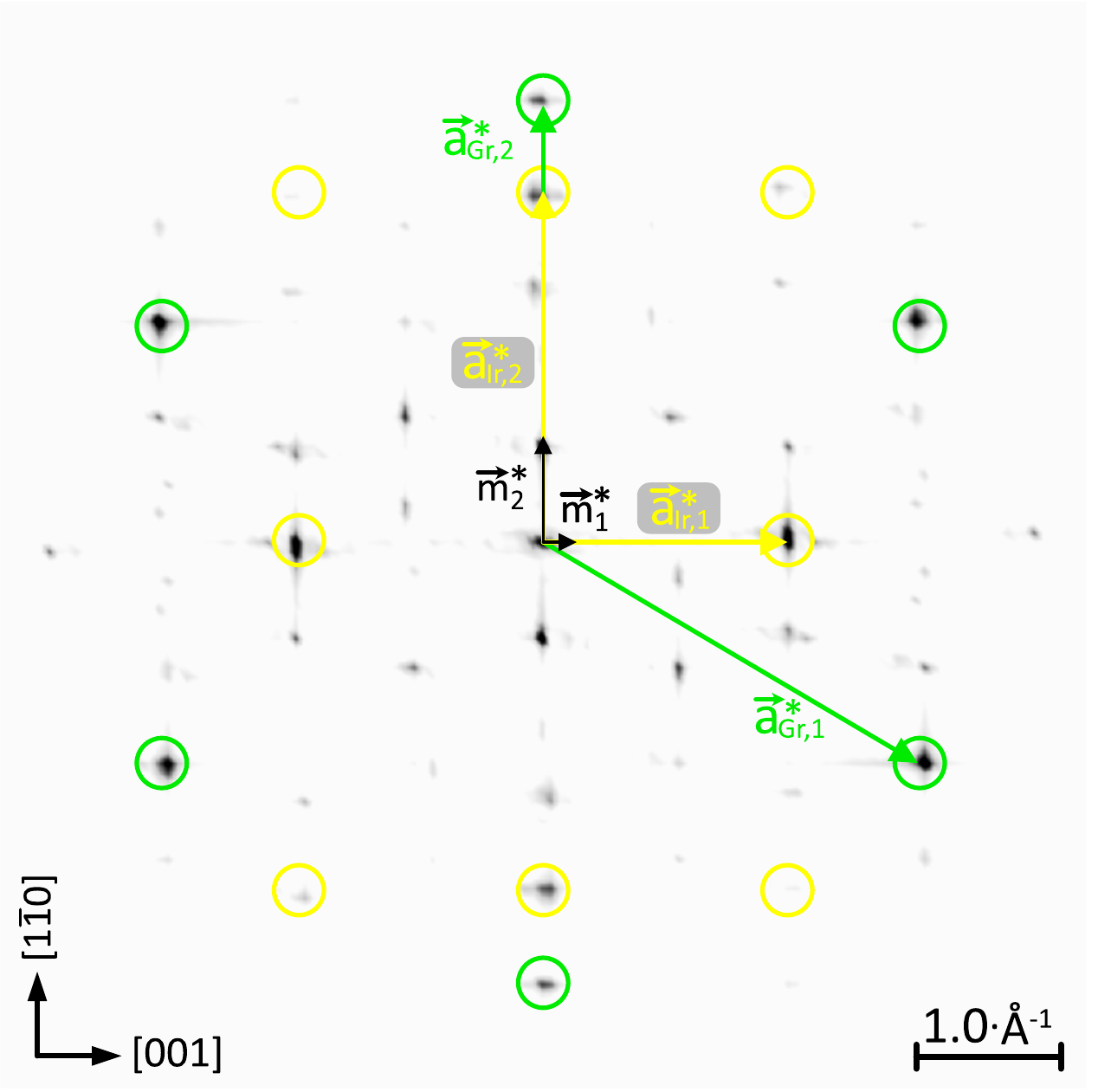}
\caption{Contrast-inverted Fourier transform of atomic resolution STM topograph of Gr/Ir(110). The reciprocal lattice points of the Ir(110) surface are encircled in yellow, the reciprocal lattice points of the single graphene domain are encircled in green. All corresponding reciprocal lattice vectors are indicated in the respective colors.}
\label{fig:Fourier}
\end{figure}

The contrast-inverted Fourier transform in Figure~\ref{fig:Fourier} shows the reciprocal lattice points of the Ir(110) substrate and the single-domain Gr. Comparing their locations allows one to conclude that Gr is undistorted, i.e. the lattice parameters $a_{\mathrm{Gr},1}$ and $a_{\mathrm{Gr},2}$ agree with each other with an error margin below $0.5$\,\%. Additionally, the reciprocal lattice points of the rectangular moiré lattice are present in the Fourier transform, and agree with the values from the STM and LEED analysis within the margin of error.

\newpage
\section*{Figure S3: Two-domain Gr on Ir(110)}

\begin{figure}[!ht]
\centering
\includegraphics[width=12cm]{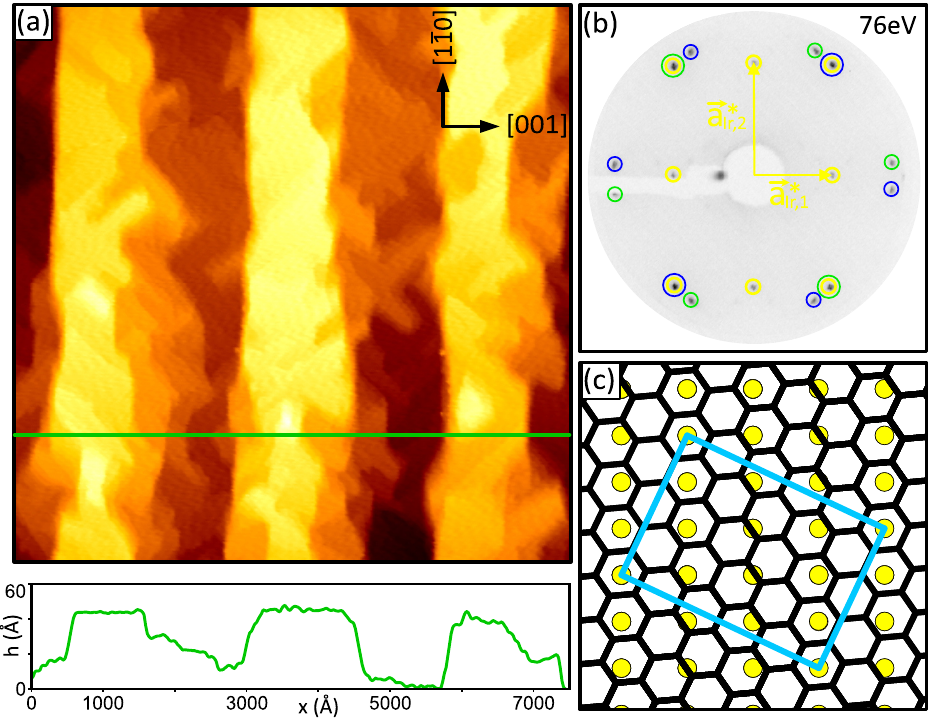}
\caption{a) STM topograph ($7500 \times 7500$\,\AA$^2$) of Gr/Ir(110) grown at $T=1300$\,K. The $[001]$ and $[1\bar{1}0]$ directions specified in upper right corner are valid also for (c). STM height profile along green line in is shown below the topograph. (b) Contrast-inverted $76$\,eV LEED pattern of sample in (a). The first order Ir reflections are encircled yellow, the first order Gr reflections related to the two Gr domains encircled green and blue, respectively. (c) Ball model of Gr domain on Ir(110), corresponding to Gr reflections encircled blue in (b). A commensurate unit cell is indicated by the cyan rectangle. STM imaging temperature in (a) is $300$\,K, tunneling parameters in (a) are $U_{\mathrm{bias}}=-1.01$\,V and $I_{\mathrm{t}}=0.97$\,nA.}
\label{fig:lowT}
\end{figure}

We have also observed a two-domain Gr layer when ethylene exposure was conducted at the lower temperature of $1300$\,K. In the large scale STM topograph in Figure~\ref{fig:lowT}a taken after Gr growth, a rough surface of ill-defined plateaus along the $[1\bar{1}0]$ direction is visible. The profile of Figure~\ref{fig:lowT}a shows a height variation of the order of few nm along the $[001]$ direction. The corresponding contrast-inverted LEED pattern in Figure~\ref{fig:lowT}b displays the Ir first order reflections encircled yellow, along with two groups of six additional first order reflections due to Gr, encircled green and blue, respectively. Each group forms an approximate hexagon, as expected for the diffraction pattern of Gr. We assign each of these groups to a distinct Gr domain. For each domain, two Gr reflections coincide with either the (1,1) and (-1,-1) or with the (1,-1) and (-1,1) reflections of Ir(110). This coincidence determines the orientation and domain structure of the Gr. It implies a substantial tensile strain in Gr of about $4$\,\% along the direction of coincidence. Both other directions are also under tensile strain of about $1.6$\,\% compared to relaxed graphite. Figure \ref{fig:lowT}c shows an atomic model of the graphene domain indicated in blue in Figure \ref{fig:lowT}b, the cyan rectangle is the commensurate superstructure unit cell. From this model, the Gr lattice parameters are $2.54$\,\AA\: in two directions, and $2.48$\,\AA\: in the remaining direction. These numbers agree well with our experiment. This low growth temperature two-domain Gr layer is not well suited as a growth substrate due to its two-domain structure and the large substrate roughness that evolves during Gr growth.

\newpage
\section*{Figure S4: Matrix notation of the approximate commensurate superstructure cell}

\begin{figure}[!ht]
\centering
\includegraphics[width=11.5cm]{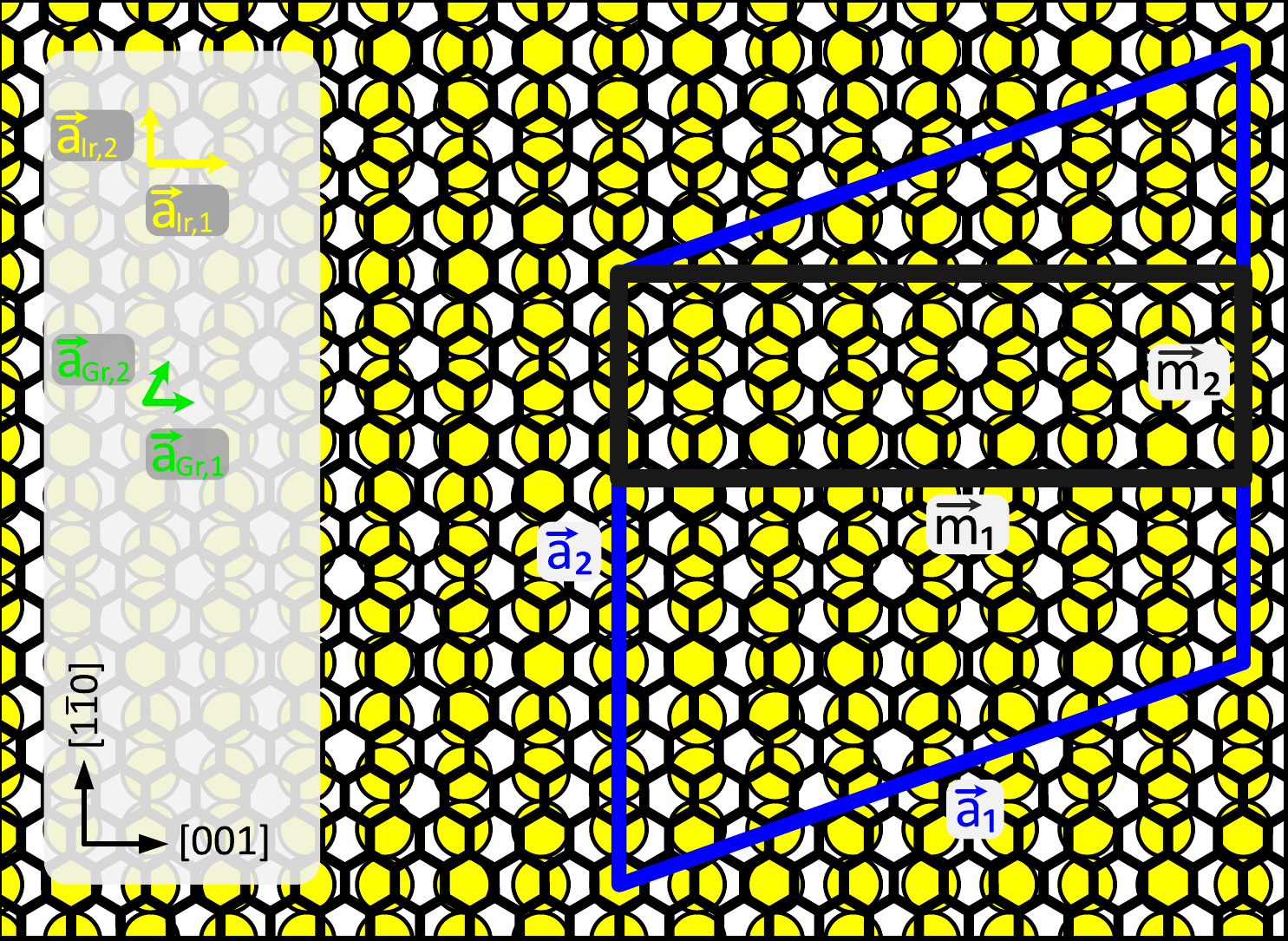}
\caption{Atomic model of the single-domain Gr phase on Ir(110). The approximate superstructure unit cell is shown as blue rhomboid, the moiré cell as black rectangle, the corresponding lattice vectors and side lengths are indicated. The Ir(110) and Gr lattice vectors are defined on the left hand side, as used for the matrix notation of the superstructure cell.}
\label{fig:supercell}
\end{figure}

Figure \ref{fig:supercell} shows an atomic model of the single-domain Gr on unreconstructed Ir(110). The superstructure and moiré cells are shown as blue rhomboid and black rectangle, respectively. Equation \ref{eq:supercell} summarizes the relation between the superstructure lattice vectors, the Ir(110) and Gr lattice vectors and is expressed in the matrix notation. All vectors are defined as shown in Figure \ref{fig:supercell}. $\vec{a}_{\mathrm{Ir},1}$ is the Ir lattice vector in $[001]$ direction, and $\vec{a}_{\mathrm{Ir},2}$ along the $[1\bar{1}0]$ direction with lengths ${a}_{\mathrm{Ir},1}=3.839$\,\AA\: and ${a}_{\mathrm{Ir},2}=2.715$\,\AA. The Gr lattice vectors in the commensurate model have lengths ${a}_{\mathrm{Gr},1}=2.457$\,\AA\: and ${a}_{\mathrm{Gr},2}=2.463$\,\AA, with $\vec{a}_{\mathrm{Gr},1}$ along $[001]$ and $\vec{a}_{\mathrm{Gr},2}$ rotated by close to $60$\,°. The resulting superstructure lattice vectors have lengths ${a}_{1}=32.58$\,\AA\: and ${a}_{2}=29.87$\,\AA.

\begin{equation}
\begin{pmatrix}
\vec{a}_1 \\
\vec{a}_2
\end{pmatrix}
=
\begin{pmatrix}
8 & 4\\
0 & 11
\end{pmatrix}
\cdot
\begin{pmatrix}
\vec{a}_{\mathrm{Ir},1} \\
\vec{a}_{\mathrm{Ir},2}
\end{pmatrix}
=
\begin{pmatrix}
10 & 5\\
-7 & 14
\end{pmatrix}
\cdot
\begin{pmatrix}
\vec{a}_{\mathrm{Gr},1} \\
\vec{a}_{\mathrm{Gr},2}
\end{pmatrix}
\label{eq:supercell}
\end{equation}

\newpage
\section*{Figure S5: Charge-density difference plots: cuts along $[1\bar{1}0]$ direction and average along [001]}

\begin{figure}[!ht]
\centering
\includegraphics[width=13cm]{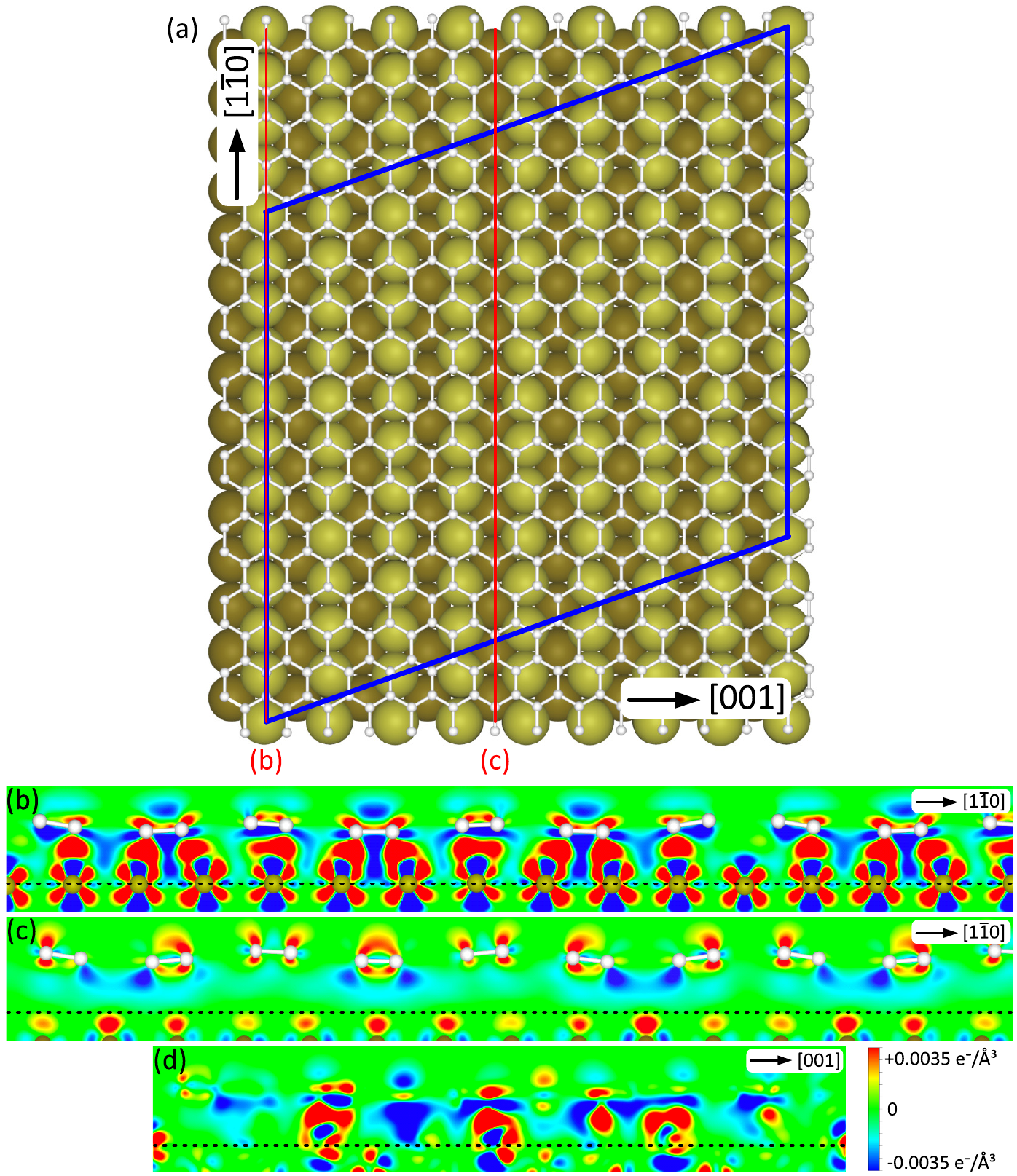}
\caption{(a) Top view of ball model representations of relaxed DFT geometries for Gr/Ir(110). Ir-atoms: light (top layer) to dark (bottom layer) brown spheres; C-atoms: small light grey dots connected by light grey lines. The DFT supercell is indicated by the blue rhomboid. (b),(c) Charge-density difference plots along $[1\bar{1}0]$-direction on top of a dense-packed Ir-row (b) and between two dense-packed rows (c), as indicated by the red lines in (a). (d) Charge-density difference average over the entire superstructure unit cell and projected onto a plane along the [001]-direction. The dotted lines in (b)-(d) indicate the vertical position of the top level Ir atoms. Color scale for (b)-(d) ranges from charge accumulation in red ($+0.0035$ electrons$/$\AA$^3$) to depletion in blue ($-0.0035$ electrons$/$\AA$^3$).}
\label{fig:DFT_charge_supplement}
\end{figure}

\newpage
\section*{Figure S6: Absence of Gr signal in ARPES on Gr/Ir(110)}

\begin{figure}[!ht]
\centering
\includegraphics[width=13cm]{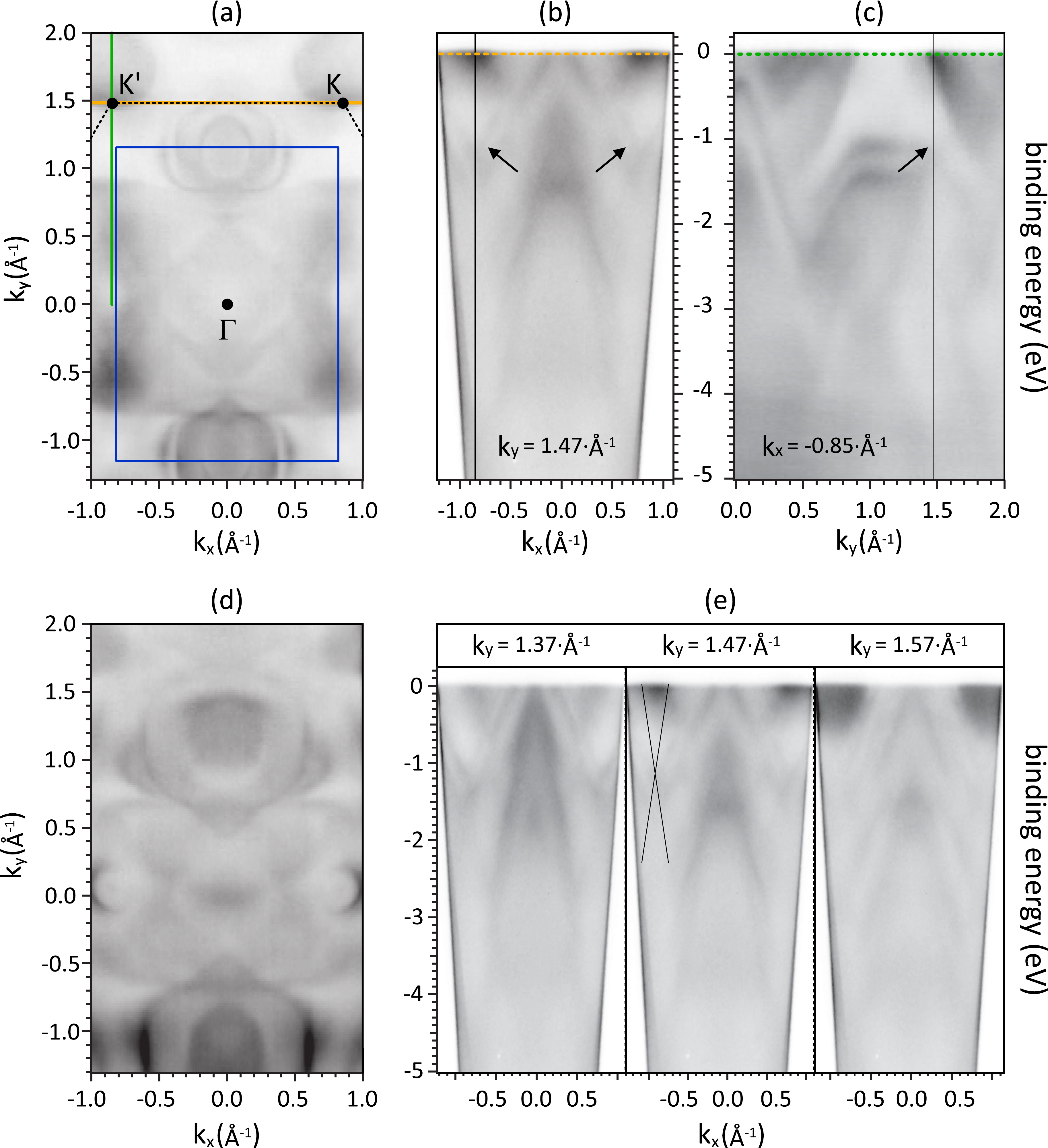}
\caption{(a) Fermi surface of Gr/Ir(110) measured by ARPES. The Ir(110) Brillouin zone is indicated by the blue rectangle, and part of the Gr Brillouin zone boundary indicated by the black dotted lines. (b) ARPES scan along $k_\mathrm{y}$ for fixed $k_\mathrm{y}=1.47$\,\AA$^{-1}$ as indicated by the orange line in (a). Expected position of Gr Dirac cone is indicated by arrows, and additionally by the vertical black line for negative $k_\mathrm{x}$. (c) ARPES scan along $k_\mathrm{y}$ with fixed $k_\mathrm{y}=-0.85$\,\AA$^{-1}$, along the green line indicated in (a). The expected position of the Gr Dirac cone indicated by vertical line and arrow. (d) ARPES constant energy slice at a binding energy of $E_{\mathrm{bind}}=-1$\,eV; no conical sections at the K points are visible. (e) ARPES scans with $\pm0.1$\,\AA$^{-1}$ larger or smaller $k_\mathrm{x}$, as compared to $k_\mathrm{y}=1.47$\,\AA$^{-1}$ used also for (b). The theoretical dispersion is shown in the center plot . All data have been recorded at a sample temperature of $T=35$\,K and a photon energy of $100$\,eV.}
\label{fig:ARPES_supplement}
\end{figure}

\newpage
\section*{Figure S7: Local adsorption minima for a naphthalene molecule on Gr/Ir(110)}

\begin{figure}[!ht]
\centering
\includegraphics[width=14cm]{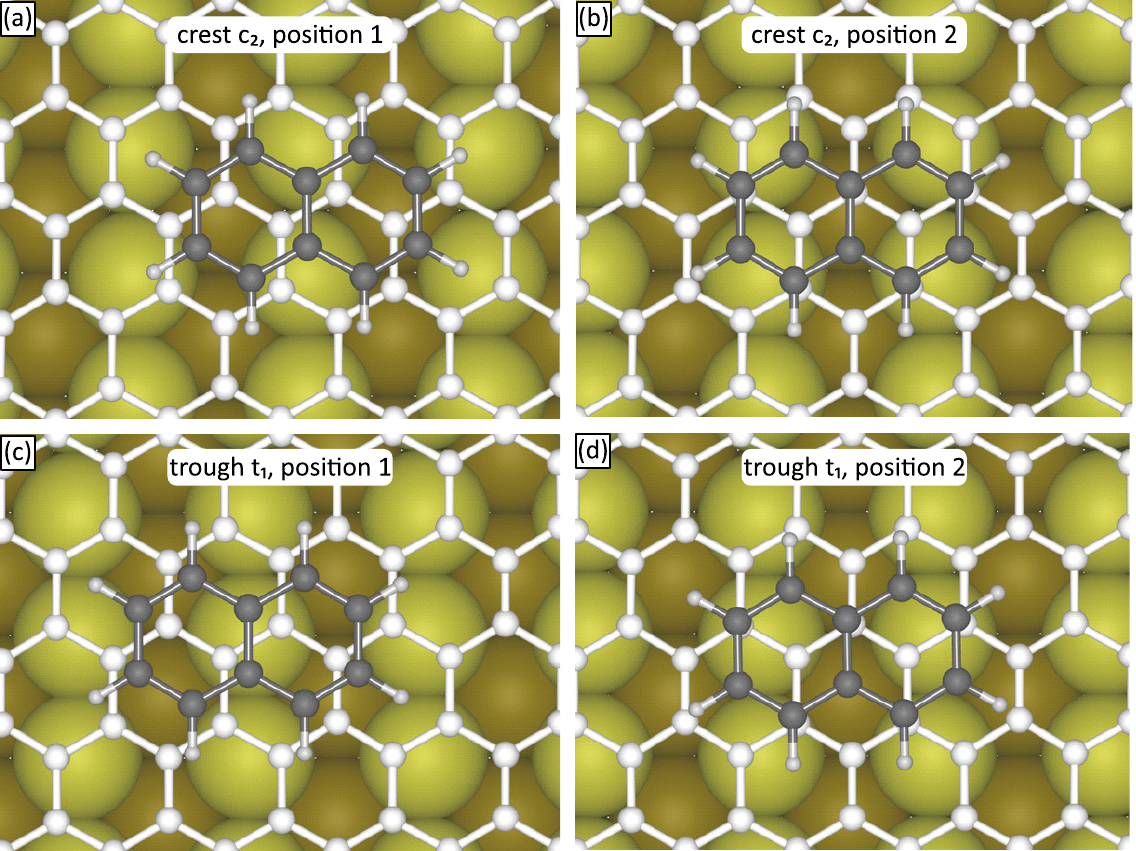}
\caption{Figures (a) to (d) display in top view ball models the local minimum energy adsorption geometries on (a),(b) crest c$_2$ and (c),(d) trough t$_1$.
The parameters of adsorption are given in Table S1.}
\label{fig:DFT_Nph_sup}
\end{figure}

Following the methodology developed in~[3], the adsorption energy $E_{\mathrm{ads}}$ can be decomposed into a term $E_{\mathrm{ads}}^{\mathrm{vdW}}$ originating from purely non-local correlation effects and the so-called DFT contribution $E_{\mathrm{ads}}^{\mathrm{DFT}}$ covering the remaining interaction. For all adsorption configurations $E_{\mathrm{ads}}^{\mathrm{vdW}}$ is negative (attactive) while $E_{\mathrm{ads}}^{\mathrm{DFT}}$ is positive (repulsive). For the minimum adsorption energy case shown in Figure~\ref{fig:DFT_Nph_sup} the numbers are $E_{\mathrm{ads}}^{\mathrm{vdW}} = -1500$\,meV per molecule ($150$\,meV per C-atom) and $E_{\mathrm{ads}}^{\mathrm{DFT}} = 553$\,meV per molecule ($55$\,meV per C-atom). This implies that binding is through purely non-local correlations, superseeding Pauli repulsion and ionic interactions by far.

\begin{table}[!h]
  \centering
    \begin{tabular}{|c|r|r|r|r|r|}
\cline{2-6}    \multicolumn{1}{r|}{} & \multicolumn{1}{c|}{E$_{\mathrm{diff}}$ (meV)} & \multicolumn{1}{c|}{E$_{\mathrm{ads}}$ (meV)} & \multicolumn{1}{c|}{E$_{\mathrm{ads}}^{\mathrm{vdW}}$ (meV)} & \multicolumn{1}{c|}{E$_{\mathrm{ads}}^{\mathrm{DFT}}$ (meV)} & \multicolumn{1}{c|}{Nph--Gr dist. (\AA)} \bigstrut\\
    \hline
    t$_1$ pos. 1 & 73.02 & -873.62 & -1389.21 & 515.60 & 3.319 \bigstrut\\
    \hline
    t$_1$ pos. 2 & 0.00  & -946.64 & -1499.66 & 553.03 & 3.272 \bigstrut\\
    \hline
    c$_2$ pos. 1 & 211.57 & -735.07 & -1202.80 & 467.72 & 3.423 \bigstrut\\
    \hline
    c$_2$ pos. 2 & 167.74 & -778.90 & -1272.93 & 464.03 & 3.392 \bigstrut\\
    \hline
    \end{tabular}
    \caption{DFT-calculated adsorption energies of naphthalene molecule adsorbed onto Gr/Ir(110). For both locations t$_1$ and c$_2$ two local minima are found (position 1 and 2). E$_{\mathrm{diff}}$ denotes the energy difference of the different local minima compared to t$_1$ (pos. 2), E$_{\mathrm{ads}}$ is the total binding energy, E$_{\mathrm{ads}}^{\mathrm{vdW}}$ is the binding energy due to non-local correlation effects and E$_{\mathrm{ads}}^{\mathrm{DFT}}$ contains the other DFT contribution to the total binding energy E$_{\mathrm{ads}}$. Nph--Gr indicates the distance in \AA\: between naphthalene and substrate.}

  \label{tab:Nph_ads}
\end{table}

\newpage
\section*{References}
\smaller[1]{[1] Koch R, Borbonus M, Haase O and Rieder K H 1991 \textit{Phys. Rev. Lett.} \textbf{67}(24) 3416–3419}
\\
{[2] Kuntze J, Speller S and Heiland W 1998 \textit{Surf. Sci.} \textbf{402-404} 764 – 769}
\\
{[3] Huttmann F, Martínez-Galera A J, Caciuc V, Atodiresei N, Schumacher S, Standop S, Hamada~I,Wehling T O, Bl\"ugel S and Michely T 2015 \textit{Phys. Rev. Lett.} \textbf{115}(23) 236101}
\end{document}